\documentclass{PoS}

\title{Validity of ChPT -- is $\;M_\pi=135\,\mathrm{MeV}\;$ small enough ?}

\ShortTitle{Validity of ChPT -- is $M_\pi$=135\,MeV small enough ?}

\author{\speaker{Stephan D\"urr}\\ 
        Wuppertal University and IAS/JSC Forschungszentrum J\"ulich\\
        E-mail: \email{\,\,durr\,(AT)\,itp.unibe.ch}}

\abstract{I discuss the practical convergence of the SU(2) ChPT series
in the meson sector, based on 2+1 flavor lattice data by the
Wuppertal-Budapest and Budapest-Marseille-Wuppertal collaborations.
These studies employ staggered and clover-improved Wilson fermions,
respectively.
In both cases large box volumes and several lattice spacings are used, and the
pion masses reach down to the physical mass point.
We conclude that LO and NLO low-energy constants can be determined with
controlled systematics, if there is sufficient data between the physical mass
point and about $350\,\mathrm{MeV}$ pion mass.
Exploratory LO+NLO+NNLO fits with a wider range reveal some distress
of the chiral series near $M_\pi\sim400\,\mathrm{MeV}$ and suggest a complete
breakdown beyond $M_\pi\sim500\,\mathrm{MeV}$.}

\FullConference{The 32nd International Symposium on Lattice Field Theory\\
                23-28 June, 2014\\
                Columbia University New York, NY}

\newcommand{\be}{\beta}

\newcommand{\ch}{\chi}

\newcommand{\bdm}{\begin{displaymath}}
\newcommand{\edm}{\end{displaymath}}
\newcommand{\bea}{\begin{eqnarray}}
\newcommand{\eea}{\end{eqnarray}}
\newcommand{\beq}{\begin{equation}}
\newcommand{\eeq}{\end{equation}}

\newcommand{\mr}{\mathrm}

\newcommand{\Nf}{N_{\!f}}

\newcommand{\Nc}{N_{\!c}}

\newcommand{\MeV}{\,\mr{MeV}}
\newcommand{\GeV}{\,\mr{GeV}}

\newcommand{\fm}{\,\mr{fm}}

\newcommand{\MSbar}{{\overline{\mr{MS}}}}

\newcommand{\Mpi}{M_\pi}
\newcommand{\Fpi}{F_\pi}
\newcommand{\Mka}{M_K}

\newcommand{\fpi}{f_\pi}

\begin{document}


\section{Introduction \label{sec:intro}}


Lattice QCD (LQCD) is the ab-initio approach to strong interactions, valid for
any value of the gauge coupling.
Chiral Perturbation Theory (ChPT) is the effective field theory approach to the
same set of phenomena, valid for small quark masses, small momenta and large box-sizes.

The chiral framework is set up as an expansion about the 2-flavor or 3-flavor
massless limit.
The SU(2) Lagrangian \cite{Gasser:1983yg} contains two low-energy constants
(LECs) at the leading order (LO), the pion decay constant $F$ and the
condensate parameter $B\!=\!\Sigma/F^2$, both defined via $m_{u,d}\!\to\!0$
and sometimes denoted $F^{(2)},B^{(2)}$, seven LECs at the NLO,
$\bar\ell_{1..7}\!\equiv\!\ln(\Lambda_{1..7}^2/[135\!\MeV]^2)$,
as well as a large number of LECs at the NNLO.
The SU(3) Lagrangian \cite{Gasser:1984gg} contains two LECs at the leading
order, $F$ and $B\!=\!\Sigma/F^2$, both defined via $m_{u,d,s}\!\to\!0$ and
sometimes denoted $F^{(3)},B^{(3)}$, ten LECs at the NLO,
$L_{1..10}^\mr{ren}(\mu\!\sim\!770\MeV)$, and a large number of LECs at the NNLO.
Below I will use $f$ and $F\!=\!f/\sqrt{2}$ in parallel, with
$\fpi^\mr{phys}\!\simeq\!130.4\MeV$ and $\Fpi^\mr{phys}\!\simeq\!92.2\MeV$.

In phenomenology, the convergence pattern is governed by the value of
$m_{ud}^\mr{phys}\simeq3.5\MeV$ [in $\MSbar$ scheme at $\mu\!=\!2\GeV$]
in the SU(2) framework, and by the value of $m_s^\mr{phys}\simeq95\MeV$
in the SU(3) framework.
The SU(2) LECs depend implicitly on $m_s^\mr{phys}$ (and heavier flavors),
and the SU(3) LECs depend implicitly on $m_c^\mr{phys}$ (and heavier flavors).

The lattice can help phenomenology by determining the numerical values of the
LECs from first principles.
Conversely, chiral formulas can aid the lattice, since they connect
different channels.
However, by using chiral formulas, one implicitly performs an extrapolation to
the respective chiral limit, and this opens the question whether the data used
are suitable to sustain that limit.
This proceedings contribution is about the selection of appropriate mass ranges
(in $\Mpi^2$ or $m_{ud}$, and possibly $2\Mka^2\!-\!\Mpi^2$ or $m_s$) to perform
the matching between the lattice data and the chiral formulas such that the
relevant LECs can be determined with controlled systematic uncertainties.


\section{Features of ChPT}




\subsection{SU(2) and SU(3) ChPT versus 2 and 2+1 and 2+1+1 flavor lattice data \label{sub:su2vssu3}}

\begin{figure}
\centering
\includegraphics[width=10cm]{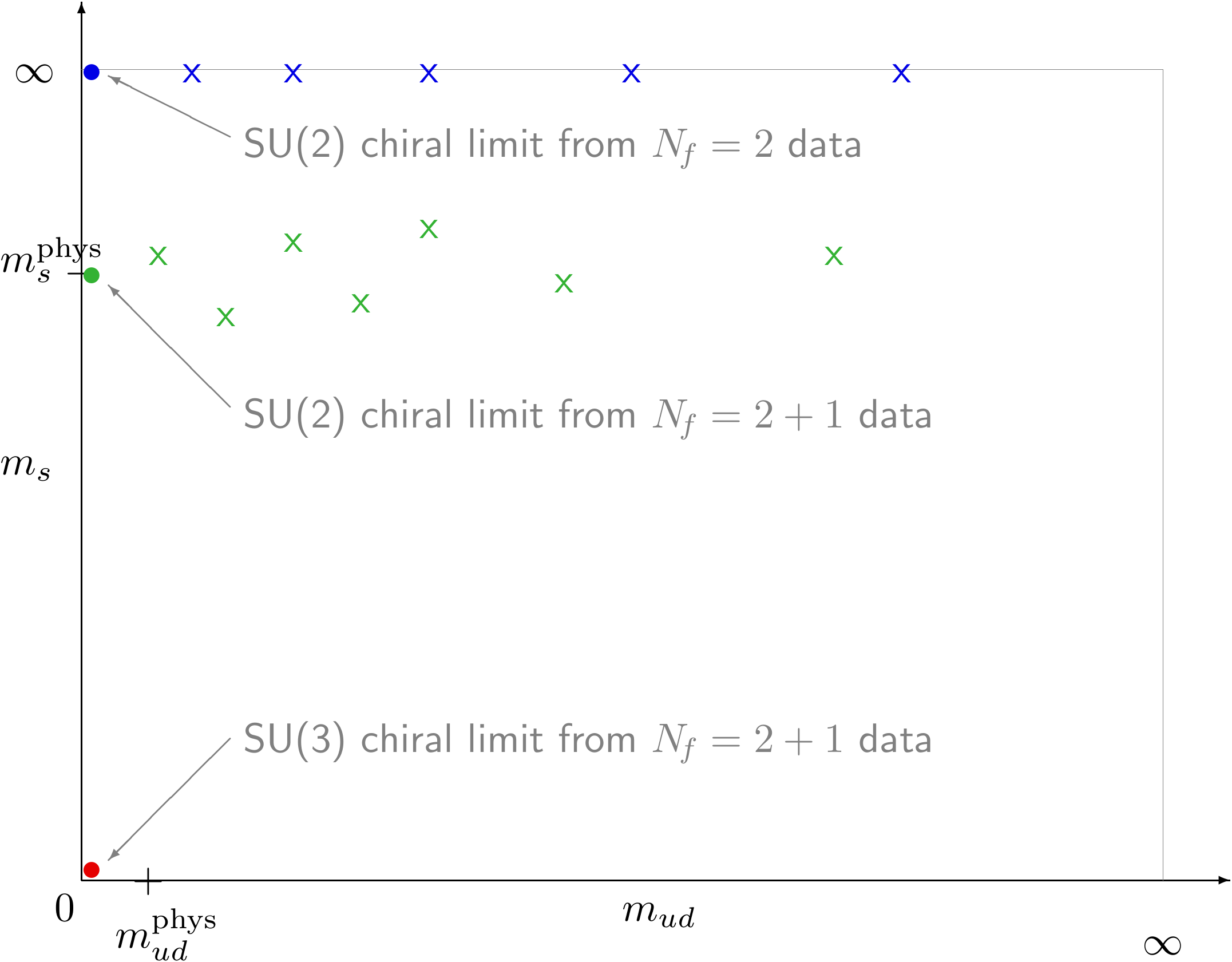}%
\caption{\label{fig:cartoon}
Sketch of different data taking strategies in the $(m_{ud},m_s)$ plane.
Simulations of QCD with $\Nf=2$ work effectively at $m_s=\infty$.
Most $\Nf=2+1$ simulations use $m_s$ values in the
vicinity of $m_s^\mr{phys}$; for a controlled extrapolation to the SU(3)
chiral limit additional data with $m_s\ll m_s^\mr{phys}$ are mandatory.}
\end{figure}

With $\Nf=2$ lattice data in hand one can only attempt to match to SU(2) ChPT.
The resulting LECs are logically different from those in phenomenology,
since they do not know about $m_s^\mr{phys}$, though the difference may be
numerically small.
With $\Nf=2+1$ or $\Nf=2+1+1$ data one has, in principle, the choice to match to
SU(2) or SU(3) ChPT.
Many collaborations opt for generating such ensembles with
$m_s\simeq m_s^\mr{phys}$, see Fig.\ref{fig:cartoon}.
In the event that there is no significant ``lever-arm'' in $m_s$, one is
restricted to comparing with SU(2) ChPT.
The advantage compared to $\Nf=2$ studies is that this time the LECs agree with
those in phenomenology (up to effects $\propto\!1/m_c^2$ or $\propto\!1/m_b^2$).


\subsection{Chiral expansion in $x$ versus in $\xi$}

Chiral formulas are often presented as an expansion in the quark mass, for instance
\bea
\Mpi^2\!&\!=\!&\!M^2\,\bigg\{1
           +\frac{1}{2}x\ln\frac{M^2}{\Lambda_3^2}
           +\frac{17}{8}x^2\Big(\!\ln\frac{M^2}{\Lambda_M^2}\Big)^2
           +x^2 k_M  +O(x^3)
\bigg\}
\label{chpt_Mpi}
\\
\Fpi\!&\!=\!&\!F\;\;\;\bigg\{1
         -x\ln\frac{M^2}{\Lambda_4^2}
         -\frac{5}{4}x^2\Big(\!\ln\frac{M^2}{\Lambda_F^2}\Big)^2
         +x^2 k_F  +O(x^3)
\bigg\} \;,
\label{chpt_Fpi}
\eea
where $x\equiv M^2/(4\pi F)^2$ with $M^2\equiv B(m_1\!+\!m_2)$.
In lattice analyses it may be convenient to invert these formulas such that they are
an expansion in $\xi\equiv \Mpi^2/(4\pi\Fpi)^2=\Mpi^2/(8\pi^2\fpi^2)$, whereupon
\bea
M^2\!&\!=\!&\!\Mpi^2\,\bigg\{1
        -\frac{1}{2}\xi\ln\frac{\Mpi^2}{\Lambda_3^2}
        -\frac{5}{8}\xi^2\Big(\!\ln\frac{\Mpi^2}{\Omega_M^2}\Big)^2
        +\xi^2 c_M  +O(\xi^3)
\bigg\}
\label{chpt_M}
\\
F\!&\!=\!&\!\Fpi\;\;\bigg\{1
      +\xi\ln\frac{\Mpi^2}{\Lambda_4^2}
      -\frac{1}{4}\xi^2\Big(\!\ln\frac{\Mpi^2}{\Omega_F^2}\Big)^2
      +\xi^2 c_F  +O(\xi^3)
\bigg\} \;.
\label{chpt_F}
\eea
The scales $\Lambda_{3,4}$ and $\Lambda_{M,F}$ or $\Omega_{M,F}$
carry no quark mass dependence (w.r.t.\ the explicitly treated flavors) and no
scale dependence.
More details, e.g.\ the relation $\Lambda_{M,F}\leftrightarrow\Omega_{M,F}$,
are found in \cite{Aoki:2013ldr}.
Some of the early discussion of the issue ``$x$ versus $\xi$ expansion'' is
found in \cite{Noaki:2008iy,Baron:2009wt,Beane:2011zm}.


\subsection{Curvature and chiral logarithms}

\begin{figure}
\includegraphics[height=53mm]{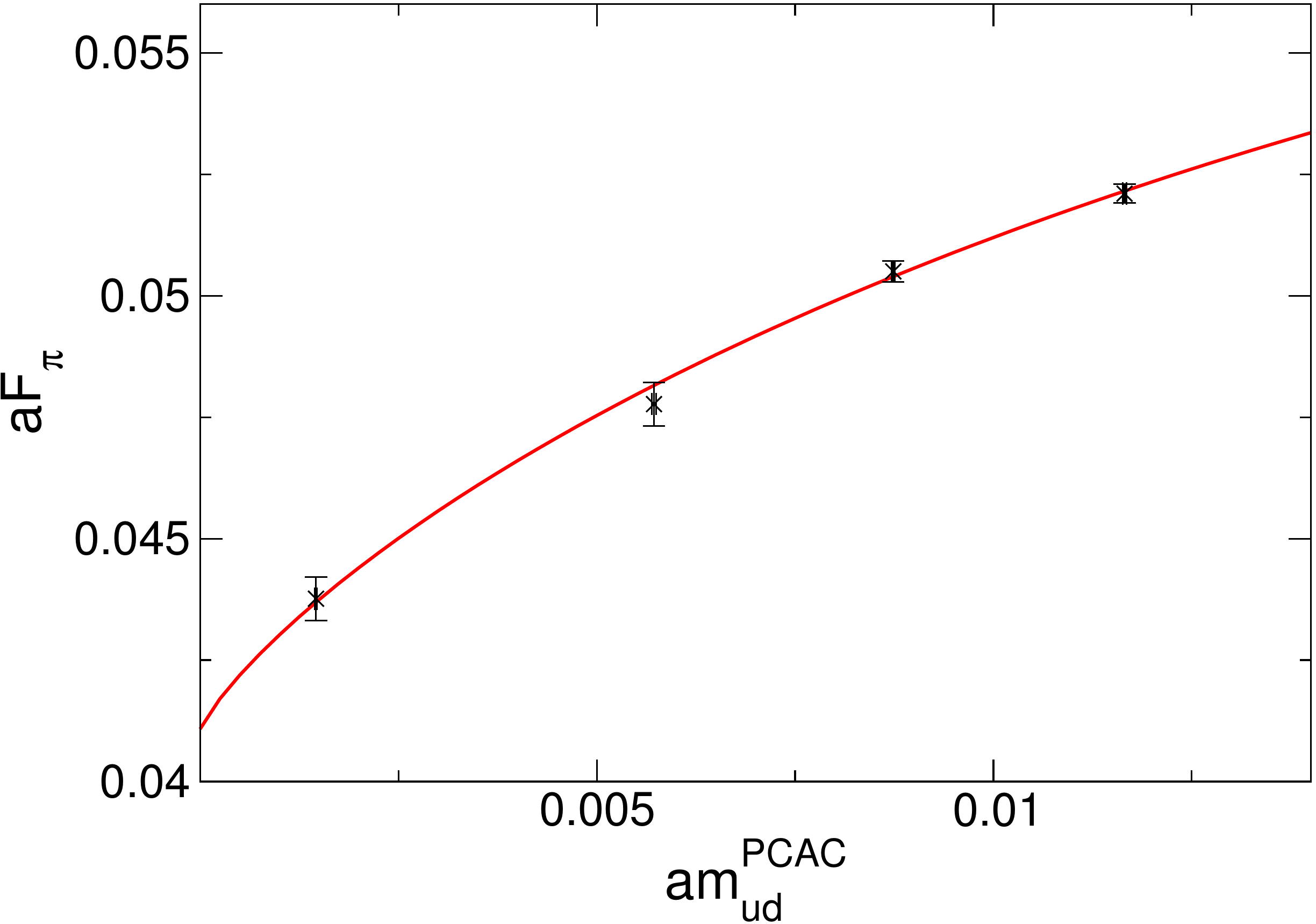}%
\includegraphics[height=54mm]{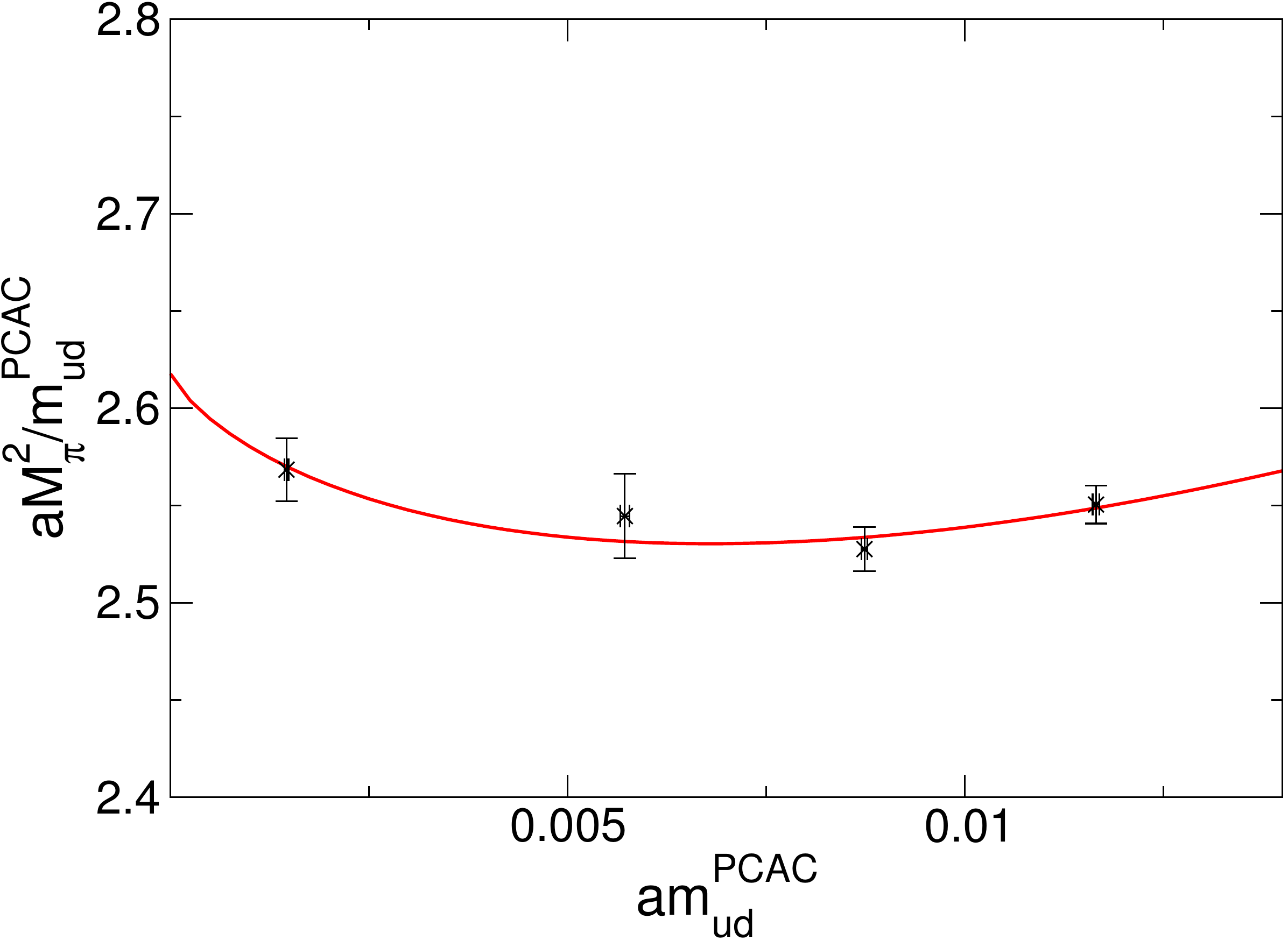}%
\vspace*{-2mm}
\caption{\label{fig:curvature}
Typical example of possible curvature in $a\Fpi$ and $a\Mpi^2/m_{ud}$ versus
$am_{ud}$. Figure taken from \cite{Durr:2010aw}.}
\end{figure}

Evidently, chiral logs are linked to the curvature inherent in data that
follow the chiral prediction (\ref{chpt_Mpi},\,\ref{chpt_Fpi}), see
e.g.\ Fig.\,\ref{fig:curvature}.
Naively, one might guess that the location of the curvature in a standard
chiral logarithm $f(M^2)=M^2\log(M^2/\Lambda_i^2)$ is linked to the scale
$\Lambda_i$.
However, taking derivatives yields $f'(M^2)=\log(M^2/\Lambda_i^2)+1$ and hence
$f''(M^2)=1/M^2$.
In other words, the curvature grows monotonically towards the chiral limit, and
this means that one needs data sufficiently close to the chiral limit to be
able to discriminate $f''(M^2)$ against zero (with the given statistics).


\subsection{Warning about finite volume effects}

\begin{figure}
\centering
\includegraphics[width=88mm]{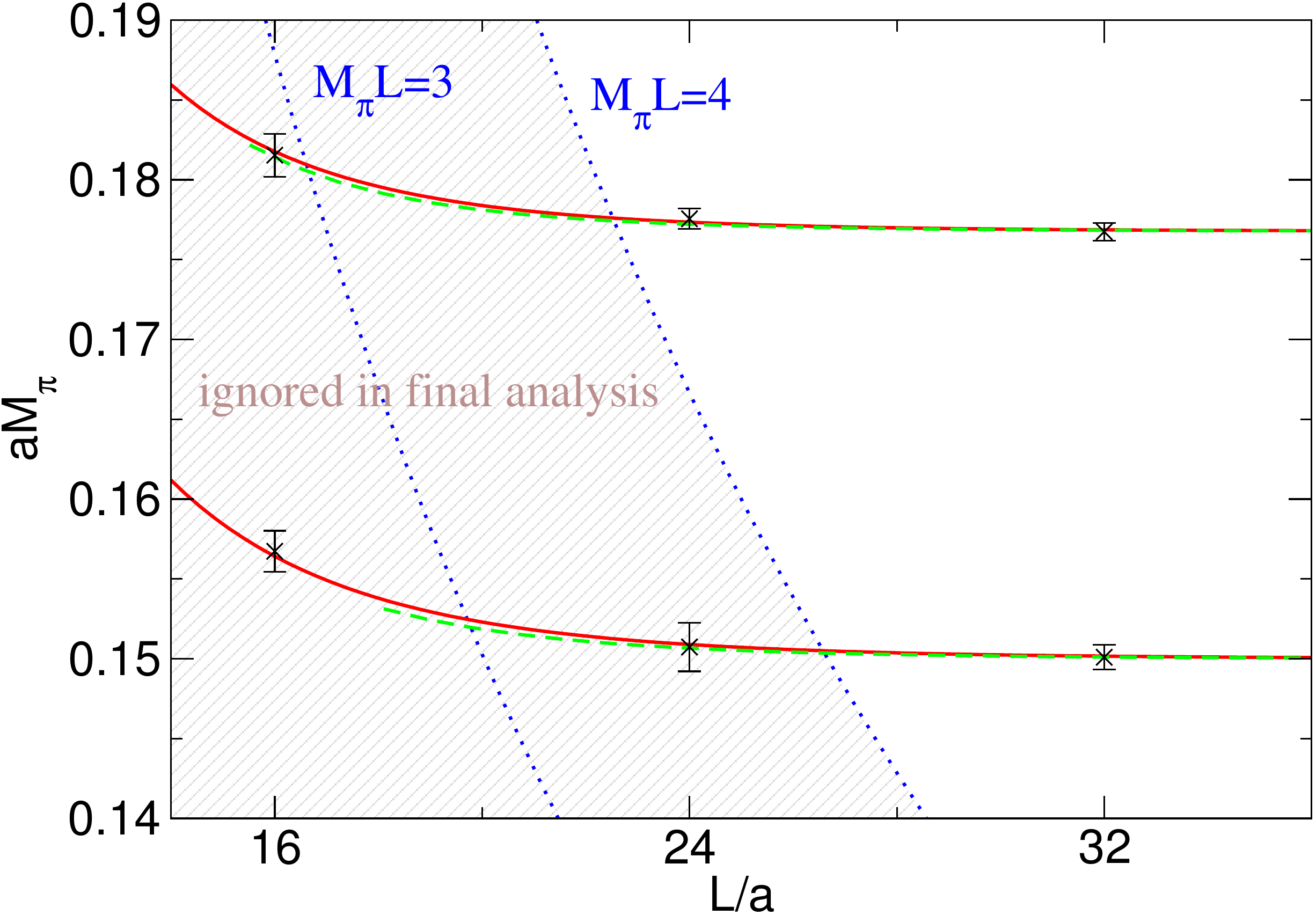}%
\vspace*{-2mm}
\caption{\label{fig:fve}
Illustration of $a\Mpi(L)>a\Mpi$ for two values of $am_{ud}$ at fixed $\be$.
Figure taken from \cite{Durr:2010aw}, in which only the three data-points with
$\Mpi(L)L>4$ would enter the final analysis.}
\end{figure}

In LQCD we work in euclidean boxes $L^3\!\times\!T$, and the finite spatial extent
$L$ creates a potential threat to the chiral expansion.
With periodic boundary conditions the lowest non-trivial momentum is $p_\mr{min}=2\pi/L$.
With $L=2\fm$ one has $p_\mr{min}\simeq630\MeV$, which is likely too much.

We are predominantly interested in the $p$-regime where $L^{-1}\ll\Mpi\ll 4\pi\Fpi$,
and the counting rule reads $m_q \sim \Mpi^2 \sim p^2 \sim L^{-2}$.
ChPT at the one-loop order predicts the finite volume effects \cite{Gasser:1986vb}
\bea
\Mpi(L)&=&\Mpi\,\Big\{1+\frac{1}{2\Nf}\xi\tilde{g}_1(\Mpi L)+O(\xi^2)\Big\}
\label{fve_mpi}
\\
\Fpi(L)&=&\,\Fpi\,\,\Big\{1-\frac{ \Nf}{2}\,\xi\,\tilde{g}_1(\Mpi L)+O(\xi^2)\,\Big\}
\label{fve_fpi}
\eea
where the shape function $\tilde{g}_1$ is given as an expansion in terms of a
Bessel function
\bea
\tilde{g}_1(z)&=&
\frac{24}{z}K_1(z)+
\frac{48}{\sqrt{2}z}K_1(\sqrt{2}z)+
\frac{32}{\sqrt{3}z}K_1(\sqrt{3}z)+
\frac{24}{2z}K_1(2z)+...
\\
K_1(z)&=&\sqrt{\frac{\pi}{2z}}\,
e^{-z}\,
\Big\{1+
\frac{3}{8z}-
\frac{3\cdot5}{2(8z)^2}+
\frac{3\cdot5\cdot21}{6(8z)^3}-
\frac{3\cdot5\cdot21\cdot45}{24(8z)^4}+...
\Big\}\;.
\eea


Finite-volume effects such as those shown in Fig.\,\ref{fig:fve} might grow
towards the chiral limit and might mimic chiral logs.
For a given set of data one wants to know whether some curvature remains
after the finite-volume effects have been compensated for.
The rule of thumb is that data with $\Mpi L\geq4$ and $L\geq3\fm$ can be corrected
for finite-volume effects by means of ChPT formulas.


\section{Investigation with staggered fermions \label{sec:stag}}


\begin{figure}
\includegraphics[width=0.32\textwidth]{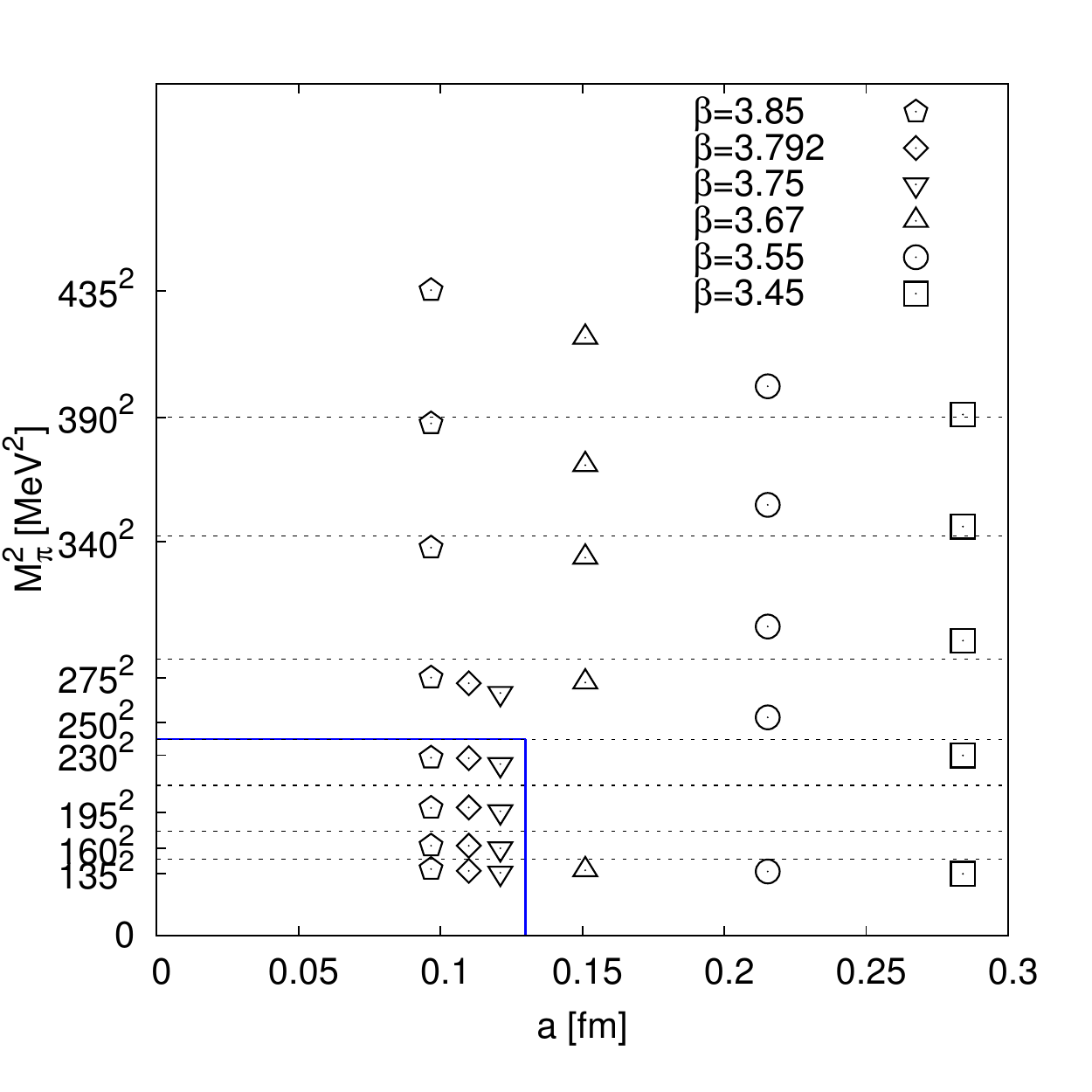}%
\includegraphics[width=0.34\textwidth]{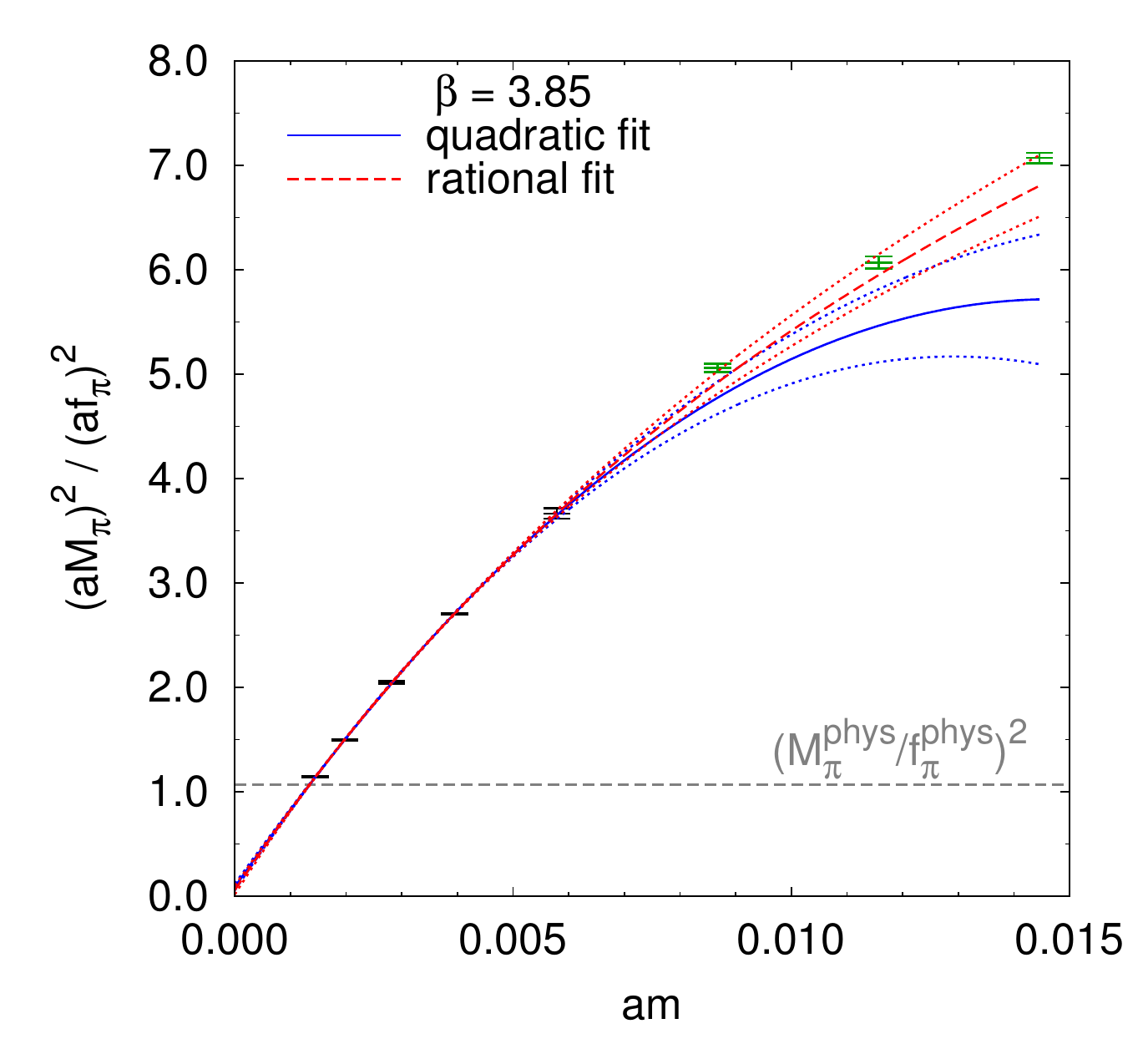}%
\includegraphics[width=0.34\textwidth]{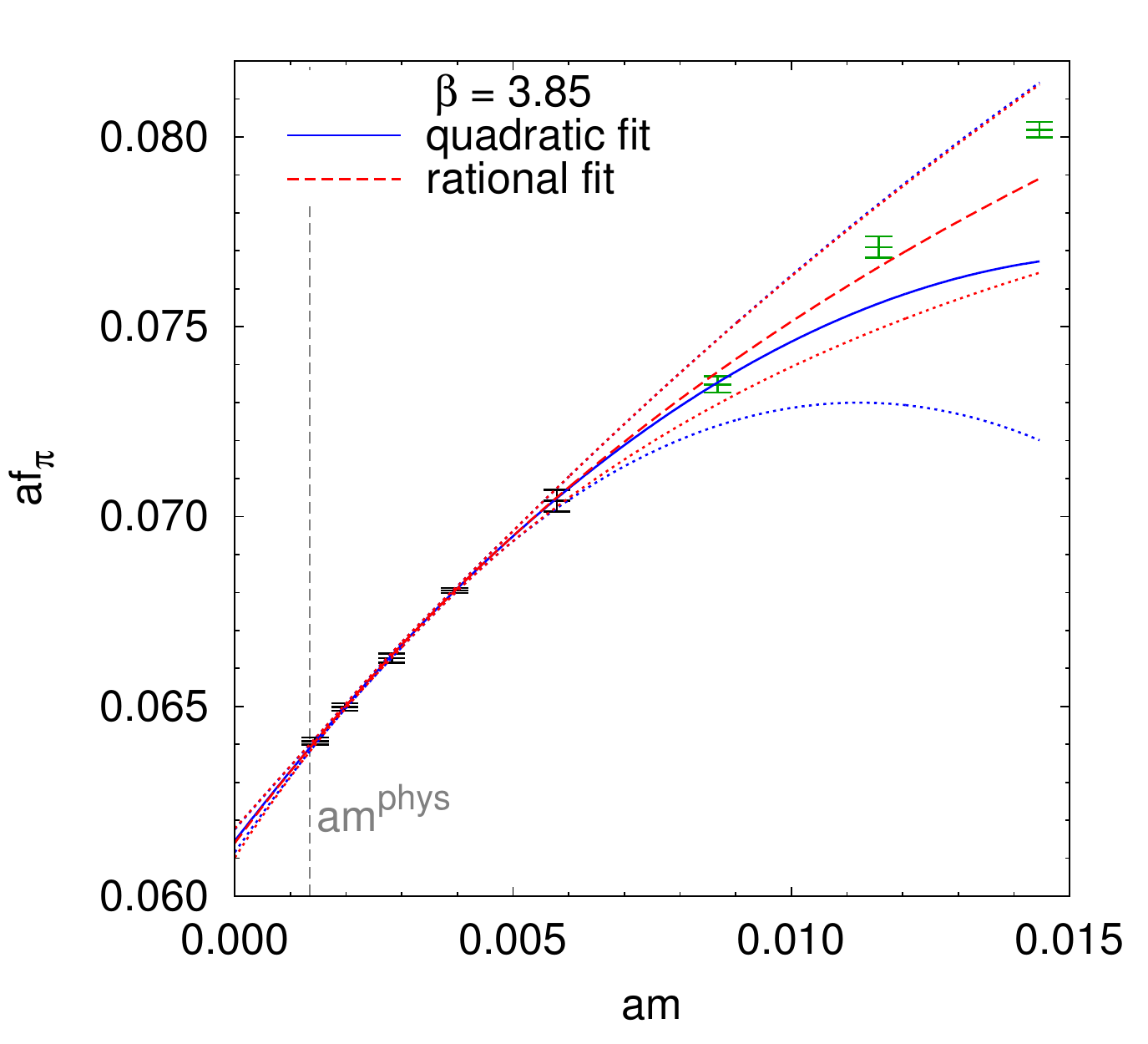}%
\vspace*{-2mm}
\caption{\label{fig:stag_landscape}
Overview of the ``landscape'' of simulation data available for the staggered
investigation (left). For each $\be$-value $\Mpi^2/\fpi^2$ is interpolated or
extrapolated, with a polynomial or rational function in $am_{ud}$, to the point
where this ratio equals 1.06846; this yields $(am)^\mr{phys}$ (middle). For
each $\be$-value the polynomial or rational function $a\fpi$ is evaluated at
$(am)^\mr{phys}$; identifying this $\fpi$ with $\fpi^\mr{PDG}$ yields $a$ (right).}
\end{figure}

The first investigation to be presented \cite{Borsanyi:2012zv} uses staggered fermions.
Fig.\,\ref{fig:stag_landscape} shows that six lattice spacings are available,
each of which features one ensemble with $\Mpi\simeq135\MeV$.
The 2nd and 3rd panel illustrate how the physical light quark mass and the
lattice spacing are determined.


\subsection{Joint SU(2) chiral fit at NLO}

\begin{figure}
\includegraphics[width=0.50\textwidth]{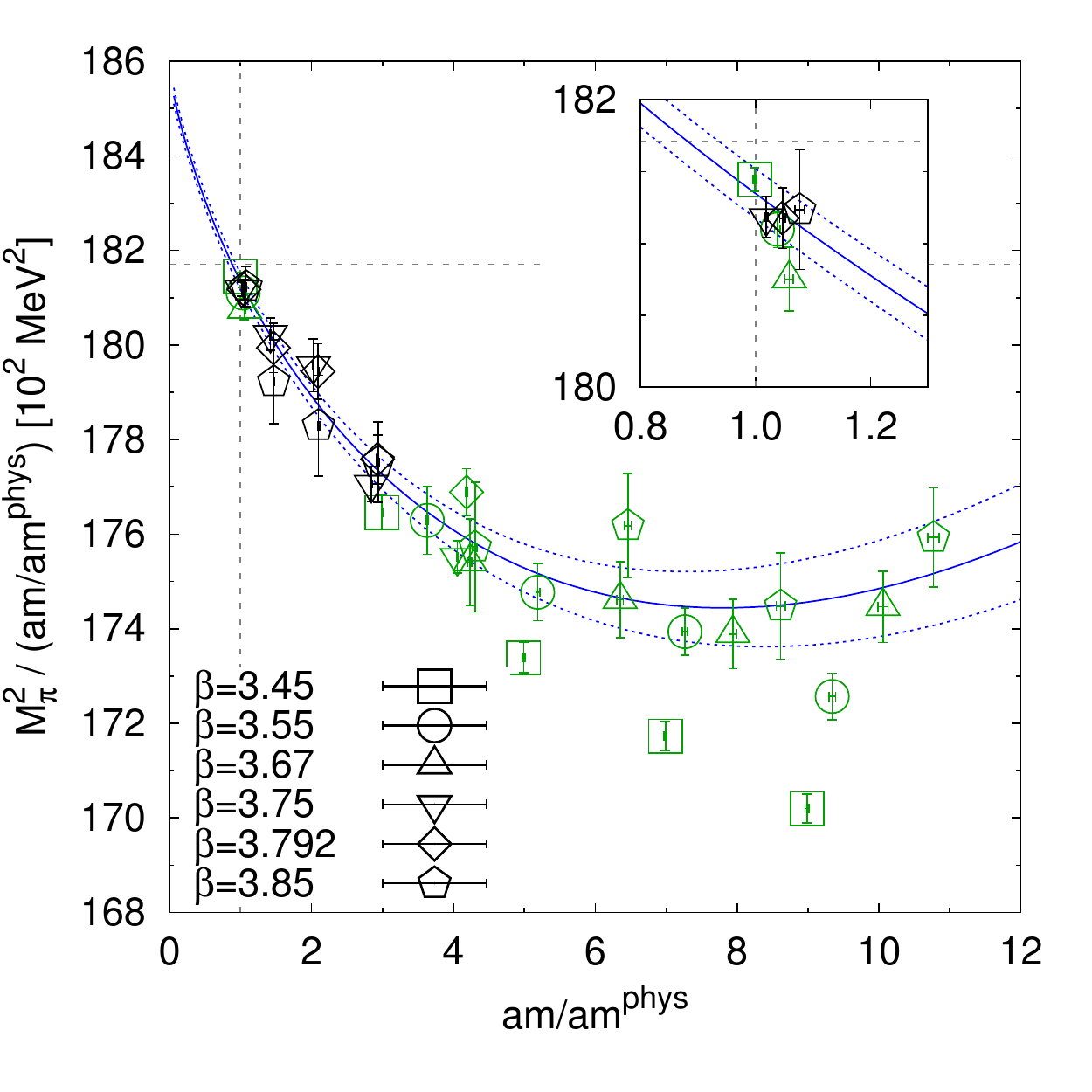}%
\includegraphics[width=0.50\textwidth]{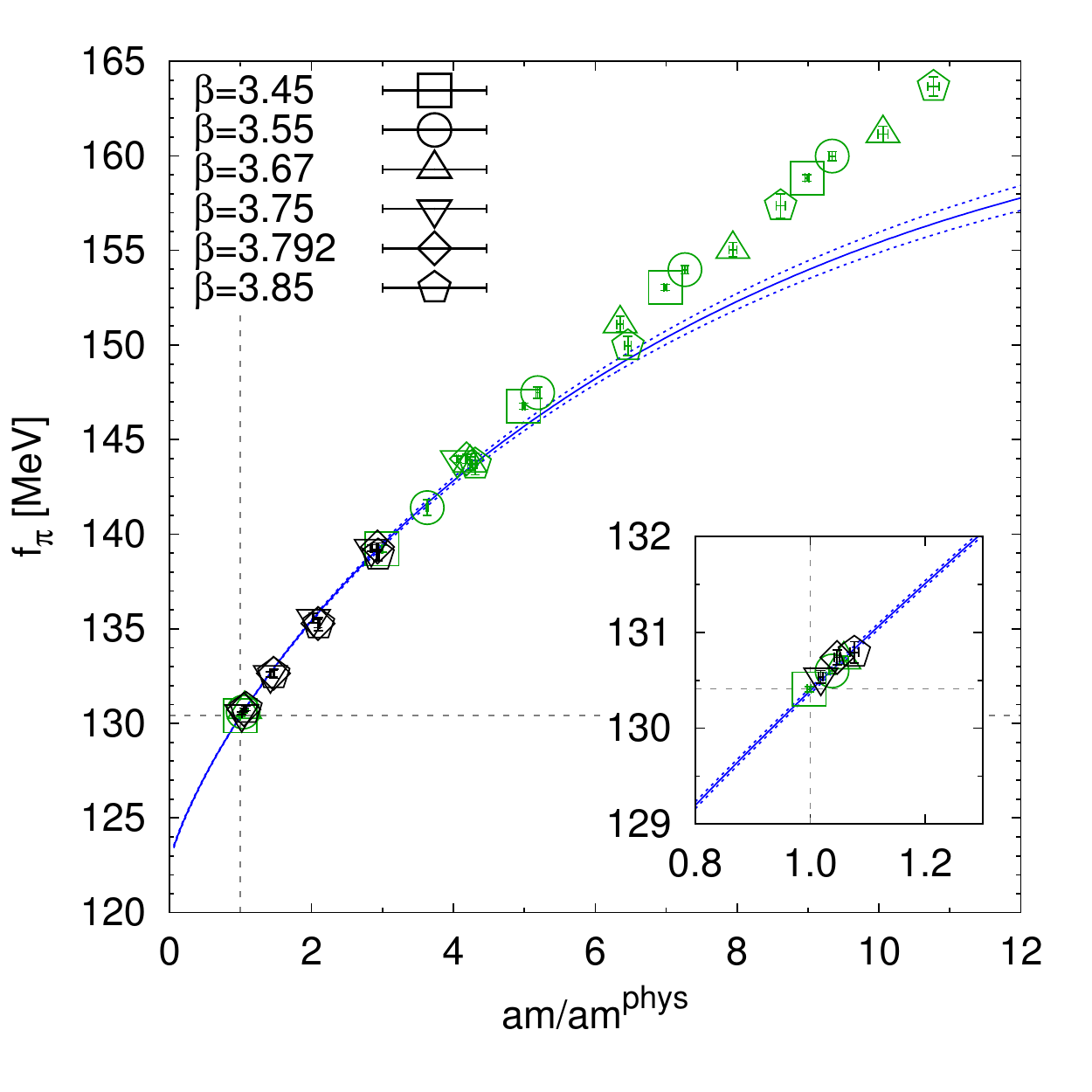}%
\vspace*{-4mm}
\caption{\label{fig:stag_preferred}
Our preferred joint LO+NLO fit to $\Mpi^2=\Mpi^2(m)$ and $\fpi=\fpi(m)$
uses data with $a^{-1}>1.6\GeV$ and $\Mpi<240\MeV$ (black data). The green
data are shown for comparison; they are not used in the fit.}
\end{figure}

The goal is to determine $\bar\ell_{3,4}$ from a joint fit to the data shown in
Fig.\,\ref{fig:stag_preferred}.
After a two-fold cut, for instance $a^{-1}\!<\!1.6\GeV$ and $\Mpi\!<\!240\MeV$,
the data can be fitted with the continuum NLO formulas
(\ref{chpt_Mpi},\,\ref{chpt_Fpi}).
The non-coarse green data deviate for $m_{ud}>5m_{ud}^\mr{phys}$ or $\Mpi>300\MeV$.


\subsection{Sensitivity of LECs on chiral range}

\begin{figure}
\includegraphics[width=0.50\textwidth]{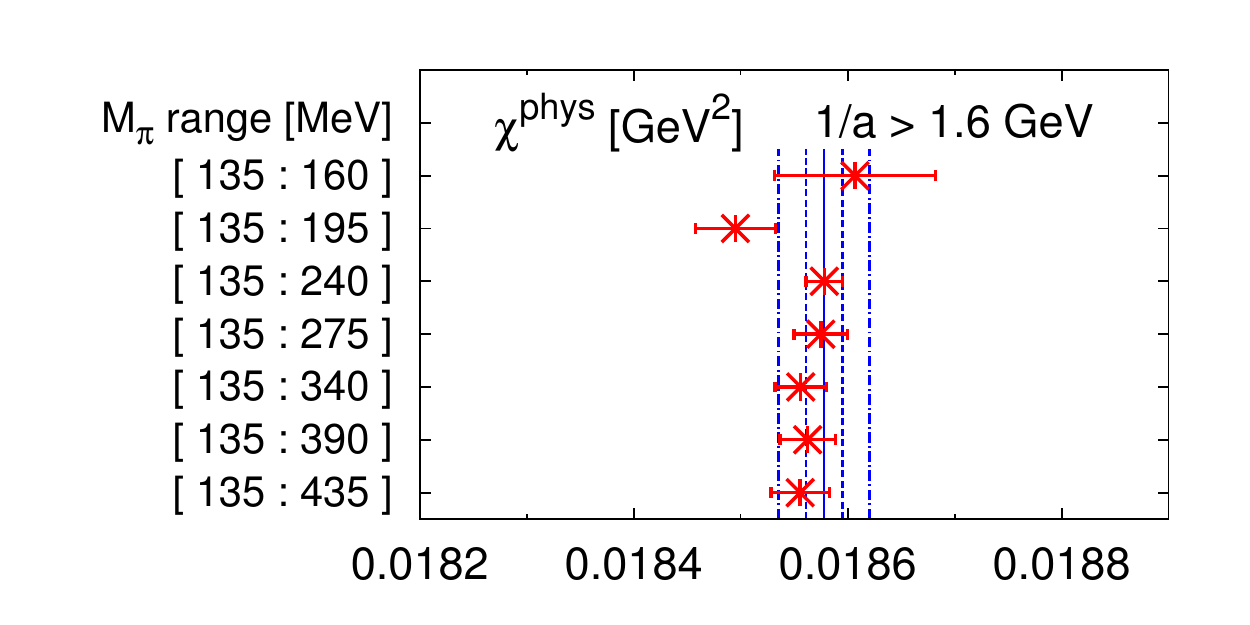}%
\includegraphics[width=0.50\textwidth]{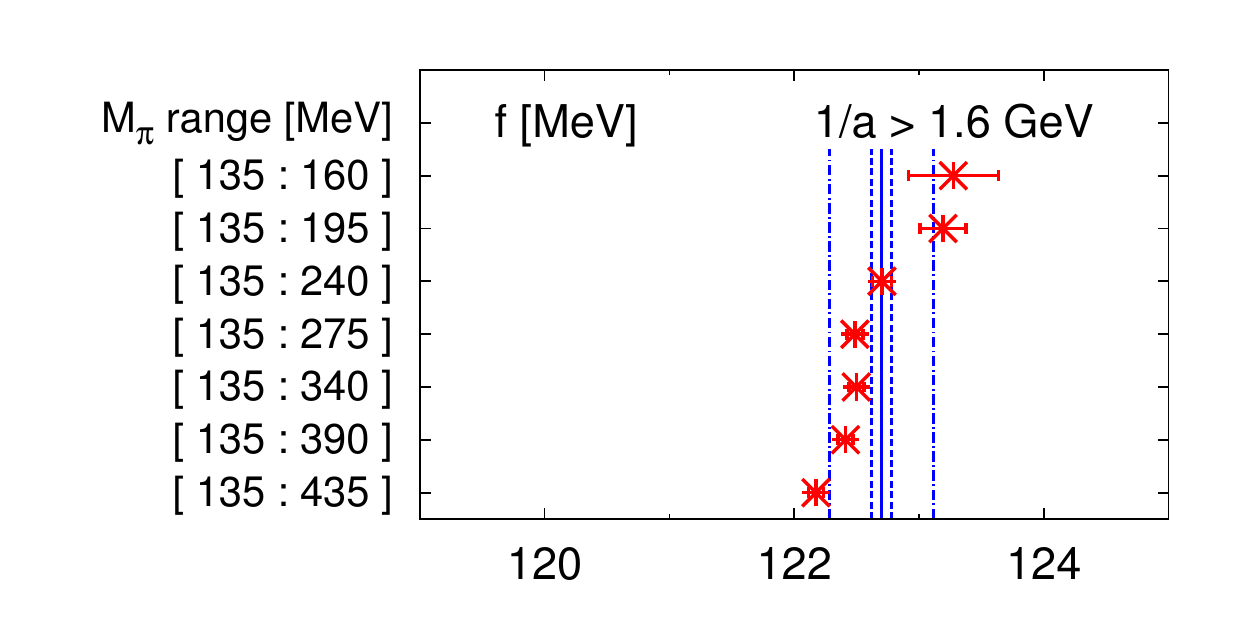}\\[-4mm]
\includegraphics[width=0.50\textwidth]{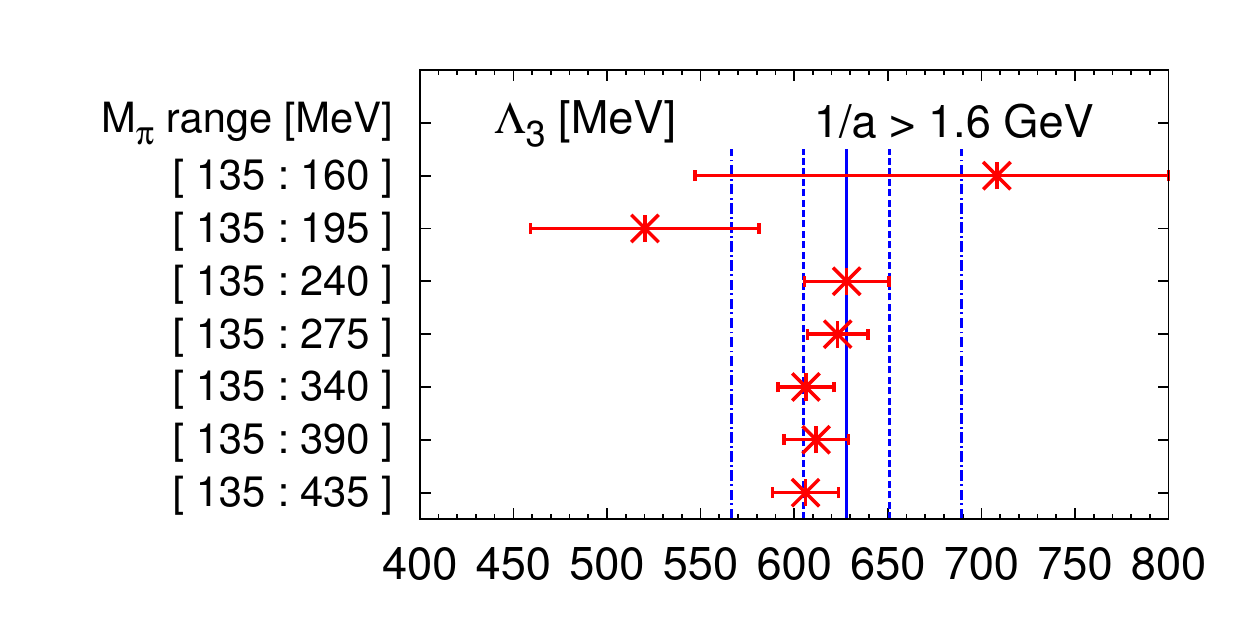}%
\includegraphics[width=0.50\textwidth]{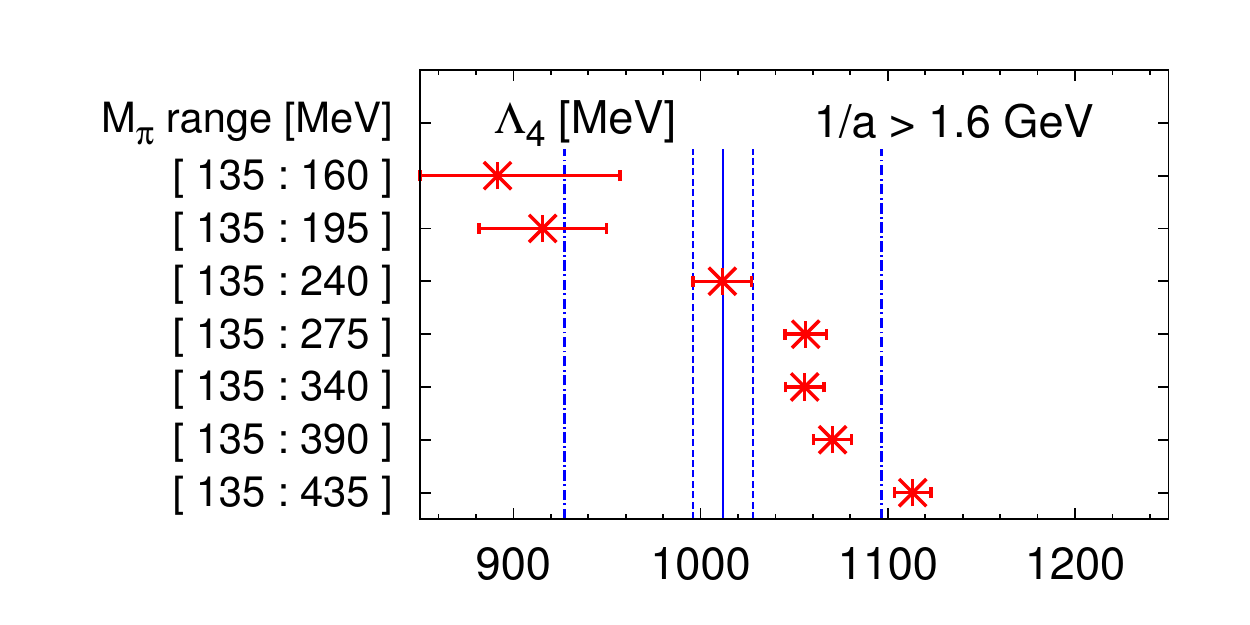}%
\vspace*{-2mm}
\caption{\label{fig:stag_results}
Summary of fitted LO ($B$ or $\ch^\mr{phys}\!\equiv\!2Bm_{ud}^\mr{phys}$, $f$)
and NLO ($\Lambda_3$, $\Lambda_4$) low-energy constants, with
$\Mpi^\mr{min}=135\MeV$ and $a^{-1}>1.6\GeV$ and several $\Mpi^\mr{max}$.
The three inner lines indicate the respective central value and statistical
error that come from the preferred fit ($\Mpi^\mr{max}=240\MeV$). The scatter
of \emph{all} results gives the systematic uncertainty which, when added in
quadrature to the statistical error, yields the outer band.}
\end{figure}

Our preferred fit features $\Mpi^\mr{max}=240\MeV$.
It is of paramount importance to compare its output to the parameters obtained
with smaller and larger $\Mpi^\mr{max}$, in order to arrive at a reliable
estimate of the theoretical uncertainty that comes from the chiral range used,
see Fig.\,\ref{fig:stag_results}.


\subsection{Sensitivity on cuts from above/below}

\begin{figure}
\includegraphics[width=0.50\textwidth]{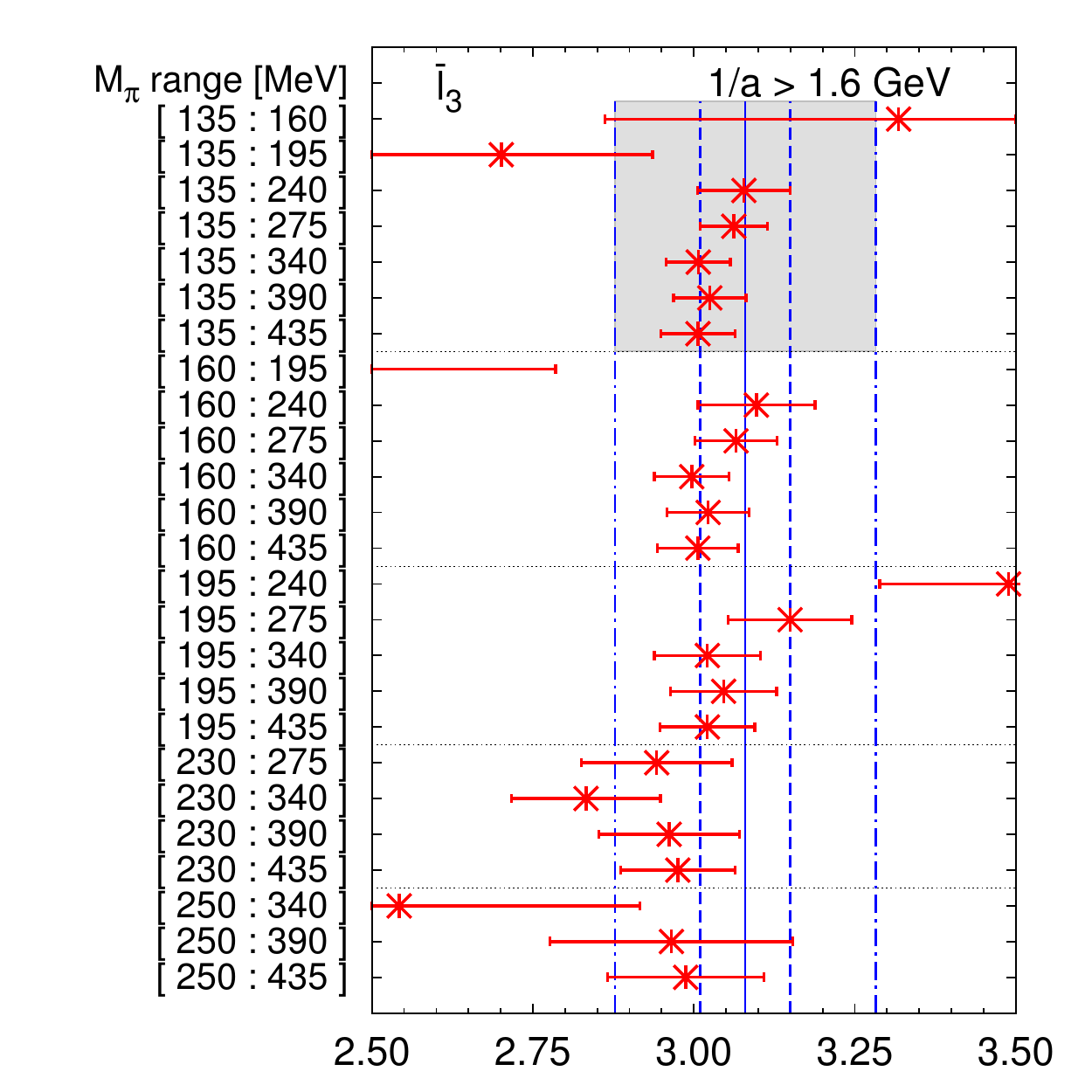}%
\includegraphics[width=0.50\textwidth]{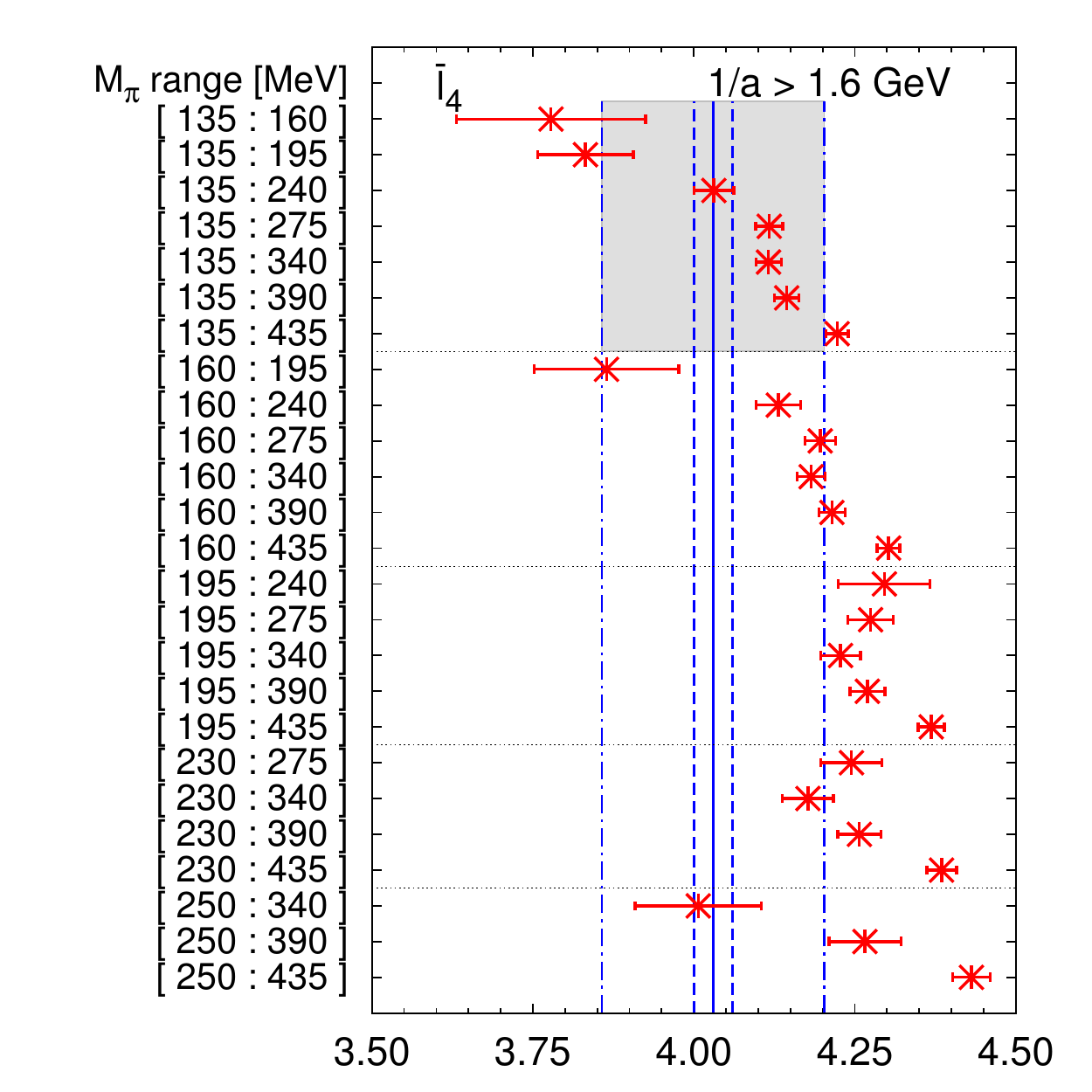}%
\vspace*{-2mm}
\caption{\label{fig:stag_extracuts}
Comparison of the NLO fits shown in the previous figure (top panel, gray
underlay) to those with $\Mpi^\mr{min}=160,195,230,250\MeV$ (lower four
panels).}
\end{figure}

One of the real benefits of a dataset that reaches down to the physical mass point
is that we can artificially prune the data from below and observe how much of
an effect this has on the fitted LO and NLO parameters.
Some of this comparison is shown in Fig.\,\ref{fig:stag_extracuts}.
Quite generally, $\ch$ and $\bar\ell_3$ are fairly robust against variations
of the chiral range, while $f$ and $\bar\ell_4$ are far more sensitive.
Choosing $\Mpi^\mr{min}$ too large tends to yield $f$ values which are too low
and $\bar\ell_4$ values which are too high.

\subsection{Breakup into LO/NLO/NNLO parts}

One of the most interesting games that one can play with such a data set is an
exploratory fit that includes the NNLO terms of equation (\ref{chpt_Mpi},\,\ref{chpt_Fpi}).
The low-energy constants $\Lambda_{M,F}$ are determined through powers of $\Lambda_{1,2}$
(which are well known from phenomenology) and $\Lambda_{3,4}$ (which are determined
through the NLO part of the fit under discussion); only $k_{M,F}$ are genuinely new.
As a result, it makes sense to include the knowledge on $\Lambda_{1,2}$ as a prior.
Still, to prevent instabilities, the fit range must be chosen somewhat wider than
in the case of the LO+NLO fit.

\begin{figure}
\includegraphics[width=0.50\textwidth]{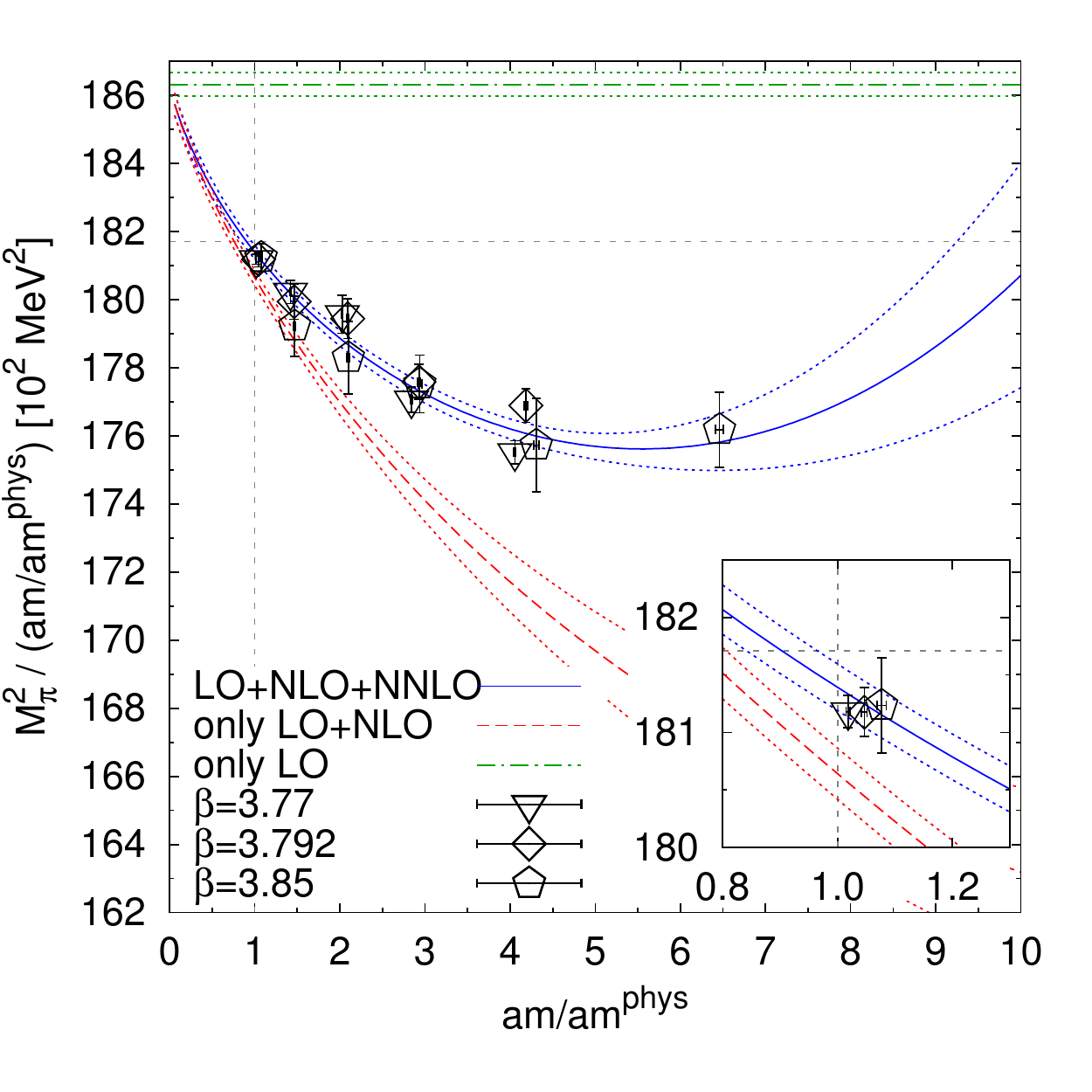}%
\includegraphics[width=0.50\textwidth]{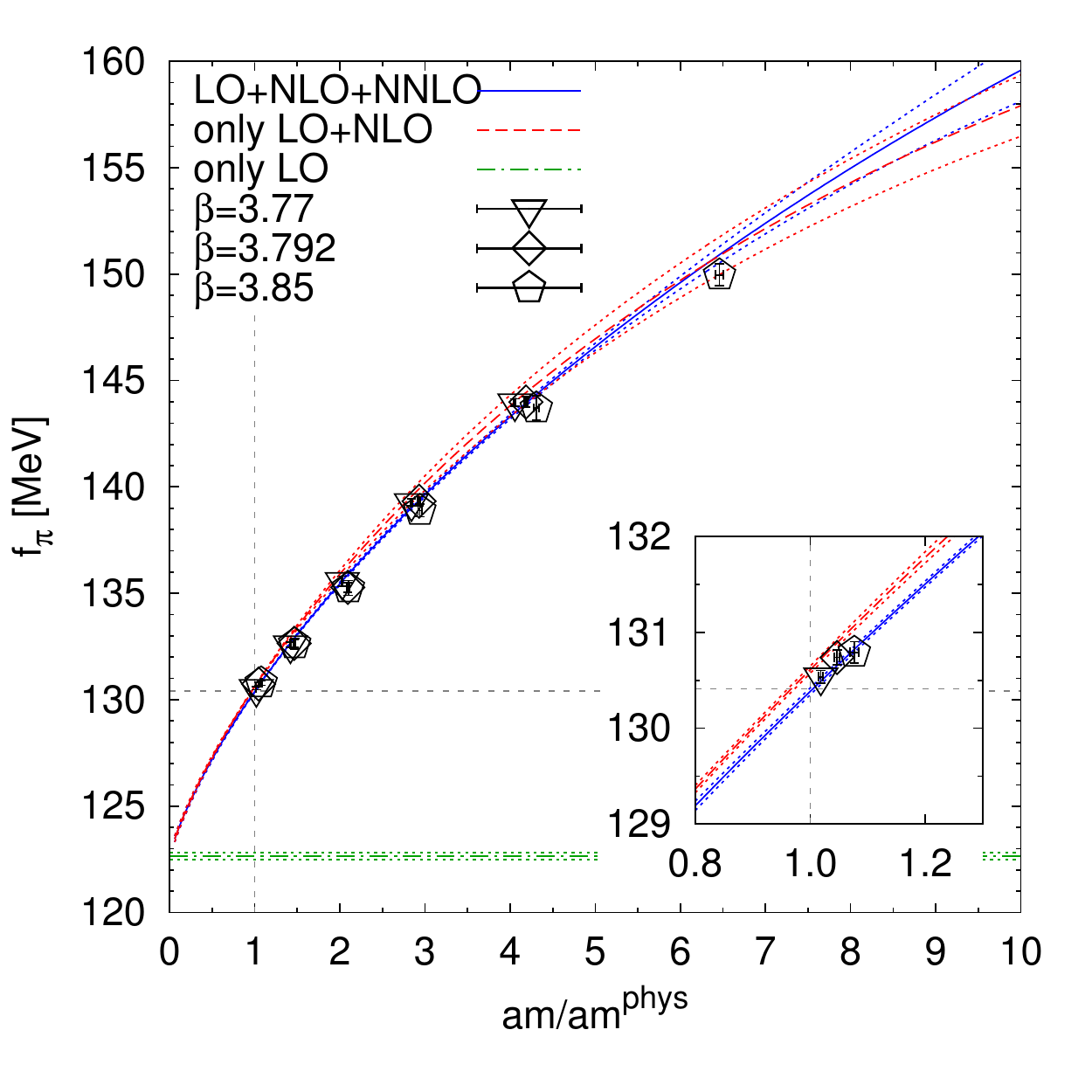}\\[-2mm]
\includegraphics[width=0.50\textwidth]{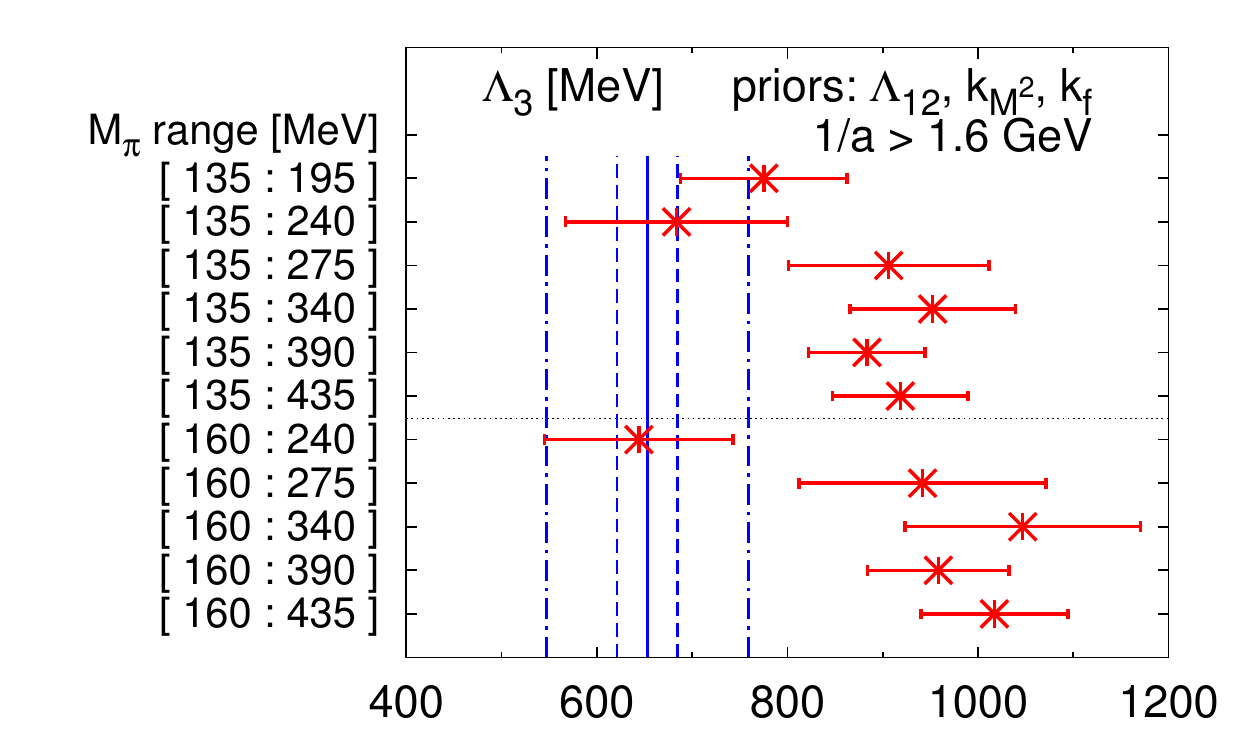}%
\includegraphics[width=0.50\textwidth]{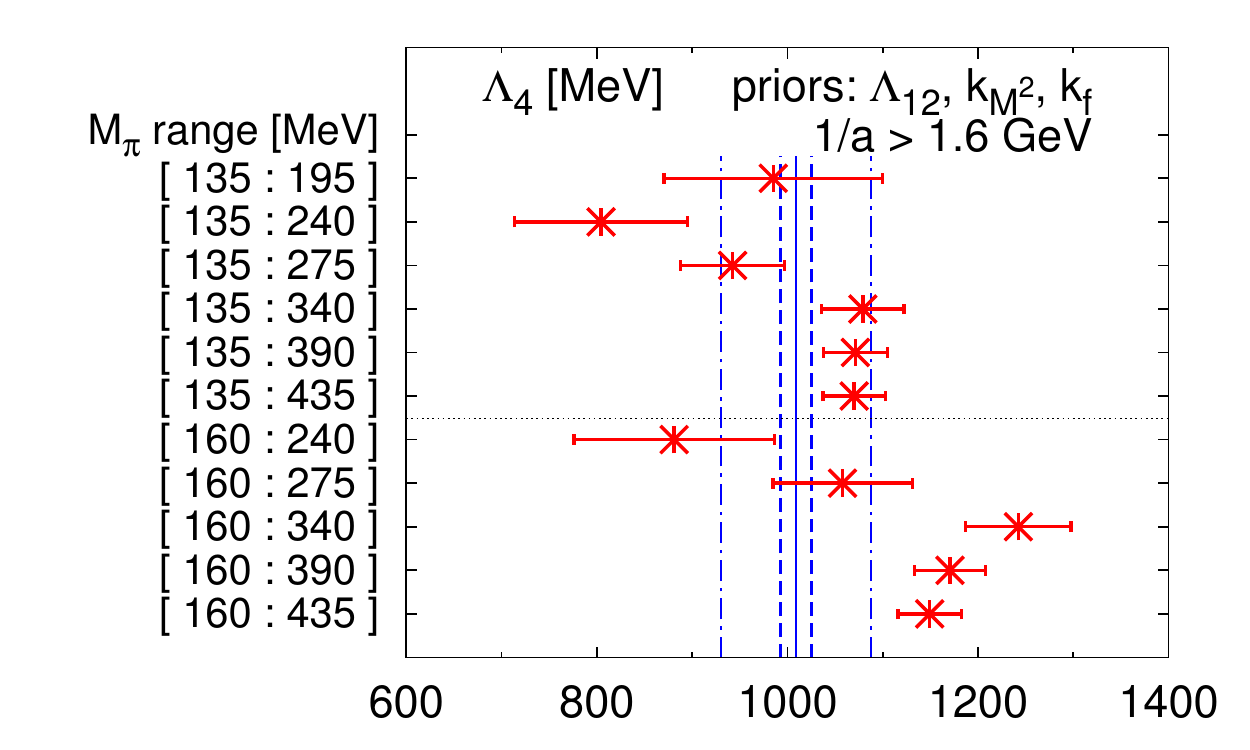}%
\vspace*{-2mm}
\caption{\label{fig:stag_exploratory}
Top: Break-up of the joint LO+NLO+NNLO fit into its LO-part (green) LO+NLO-part
(red) and the full thing (blue). Bottom: Comparing the resulting $\Lambda_{3,4}$
to those from Fig.\,6 (indicated by the blue lines).}
\end{figure}

Fig.\,\ref{fig:stag_exploratory} shows the break-up of the LO+NLO+NNLO
fit into its LO-part (green) LO+NLO-part (red) and the full thing (blue).
At $m_{ud}=m_{ud}^\mr{phys}$ we find an excellent convergence pattern, that is
$|\mr{NNLO}|\ll|\mr{NLO}|$.
At $m_{ud}=7m_{ud}^\mr{phys}$ or $\Mpi\!\sim\!350\MeV$ we find
$|\mr{NNLO}|\simeq\frac{1}{2}|\mr{NLO}|$, which marks the beginning of
some distress on the chiral series.
At $m_{ud}\simeq11m_{ud}^\mr{phys}$ or $\Mpi\!\sim\!450\MeV$ the
ordering seems to get lost, and the chiral expansion breaks down.

\section{Investigation with Wilson fermions \label{sec:wils}}


The second investigation to be presented \cite{Durr:2013goa} uses tree-level
clover improved Wilson fermions.
Again, we use only data with $m^\mr{sea}=m^\mr{val}$.
The presence of additive quark mass renormalization invites employing a global fit.
This, in turn, suggests studying the issue of $x$ versus $\xi$ expansion.


\subsection{NLO fit via $x$ and $\xi$ expansion}

\begin{figure}
\includegraphics[width=0.5\textwidth]{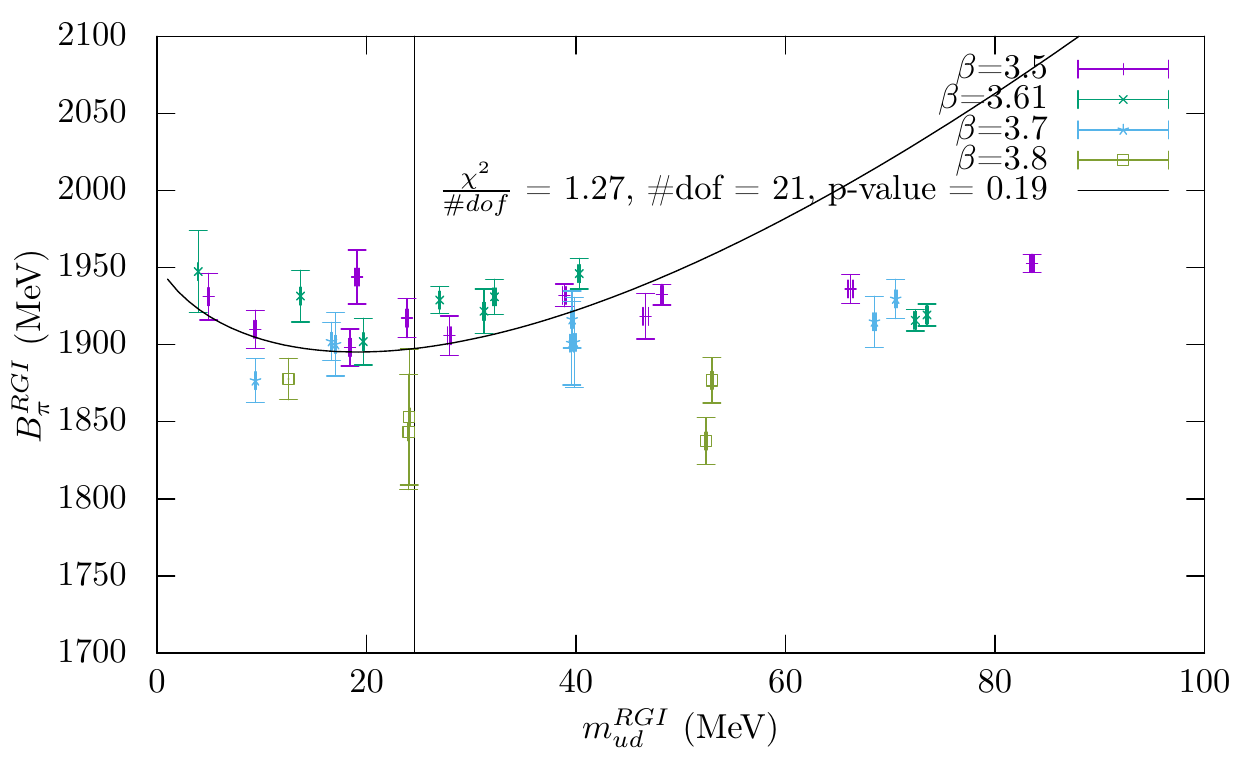}%
\includegraphics[width=0.5\textwidth]{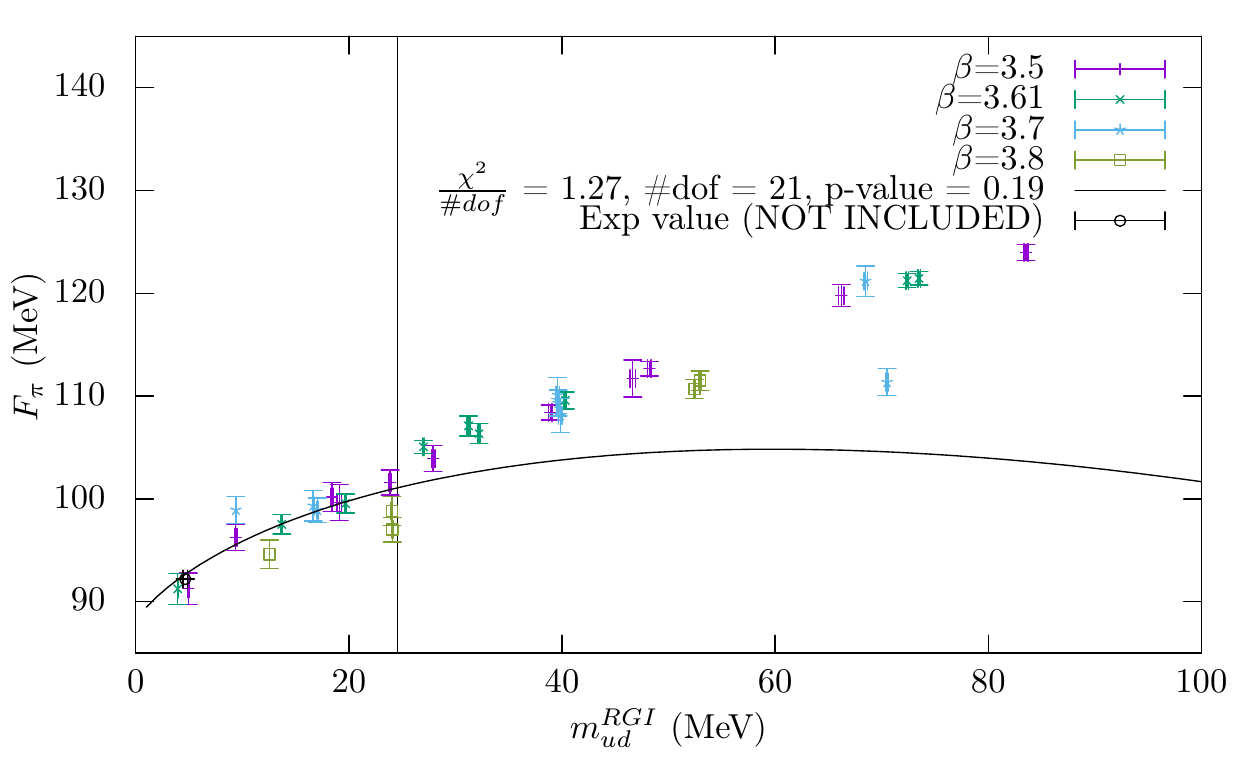}\\
\includegraphics[width=0.5\textwidth]{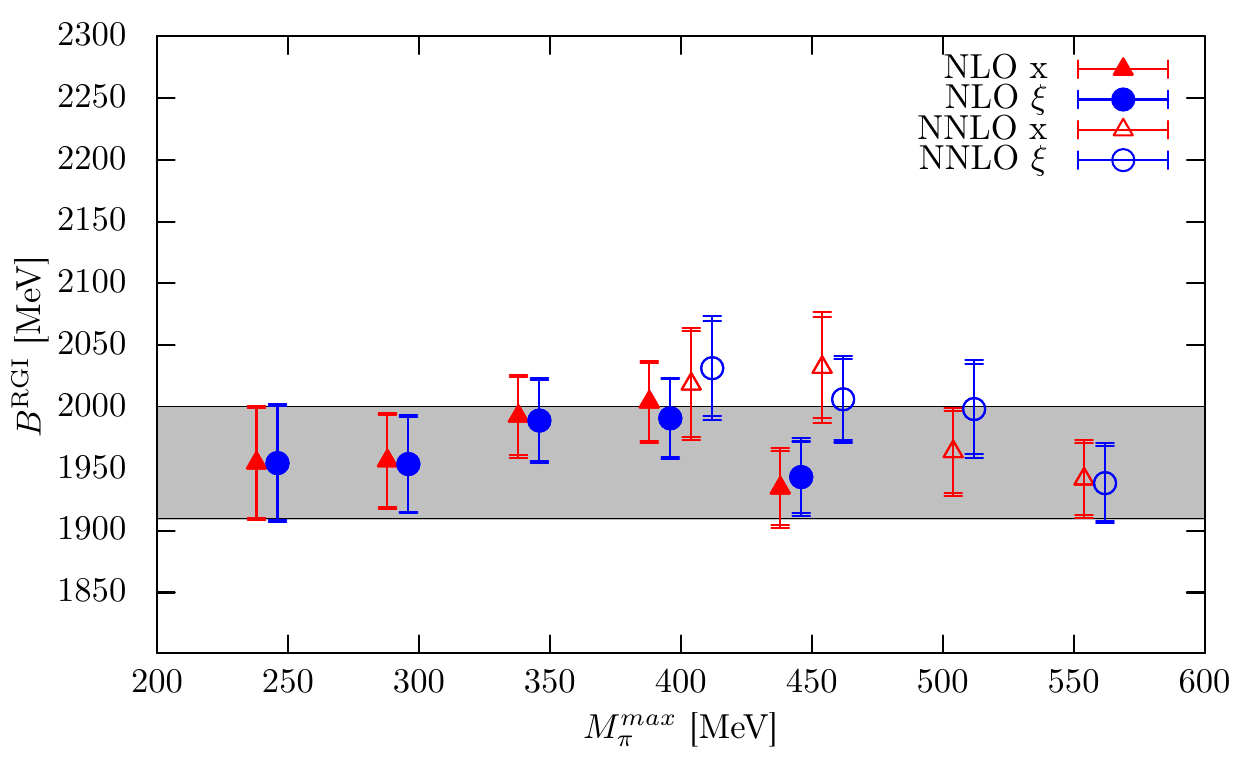}%
\includegraphics[width=0.5\textwidth]{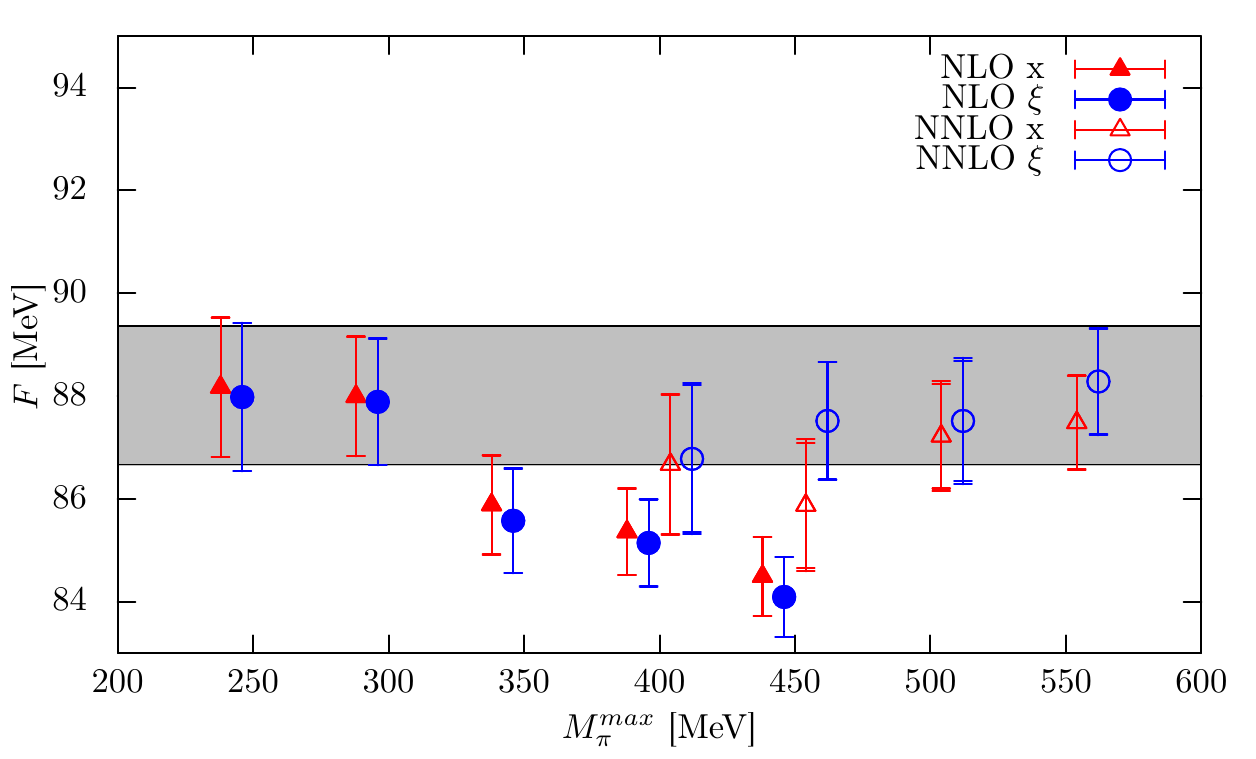}%
\vspace*{-2mm}
\caption{\label{fig:wils_NLO_x}
Top: $x$ expansion fit at NLO of $\Mpi^2/m_{ud}$ and $\Fpi$ versus $m_{ud}$
with $\Mpi^\mr{max}=300\MeV$.
Bottom: fit parameters $B,F$ as a function of $\Mpi^\mr{max}$.}
\end{figure}

\begin{figure}
\includegraphics[width=0.5\textwidth]{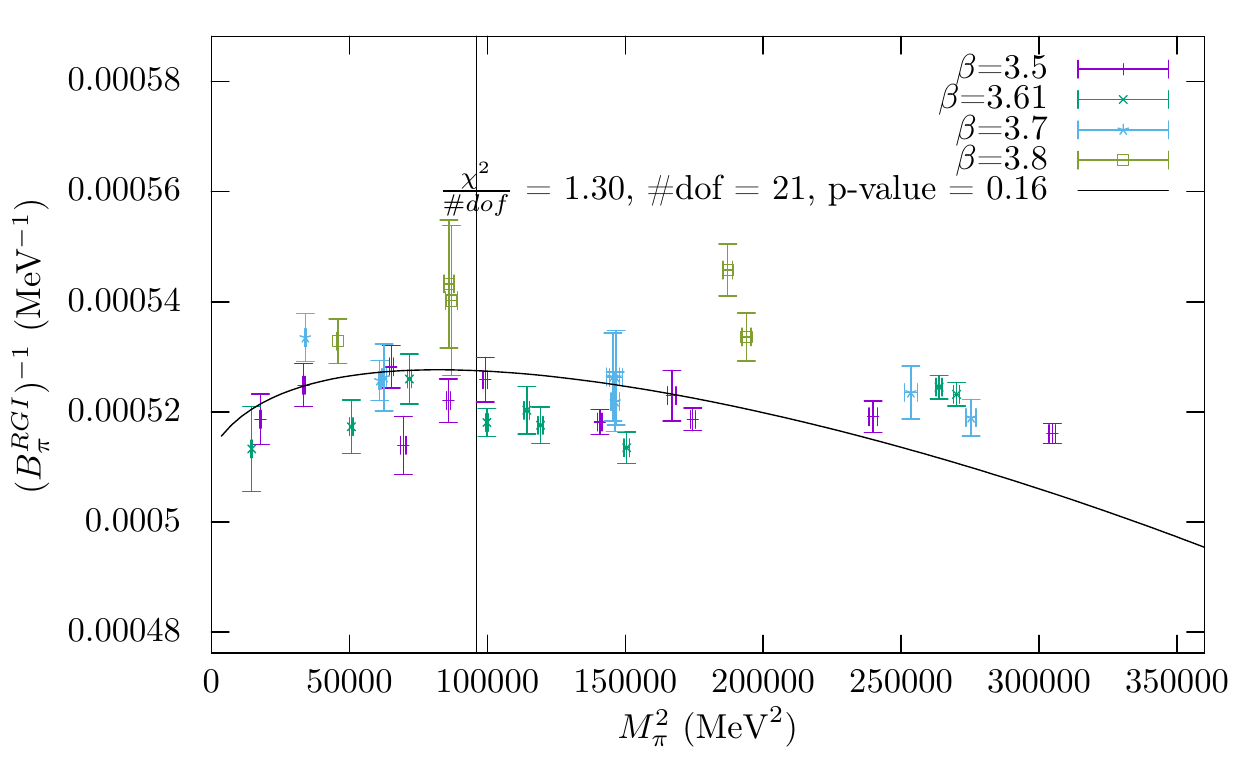}%
\includegraphics[width=0.5\textwidth]{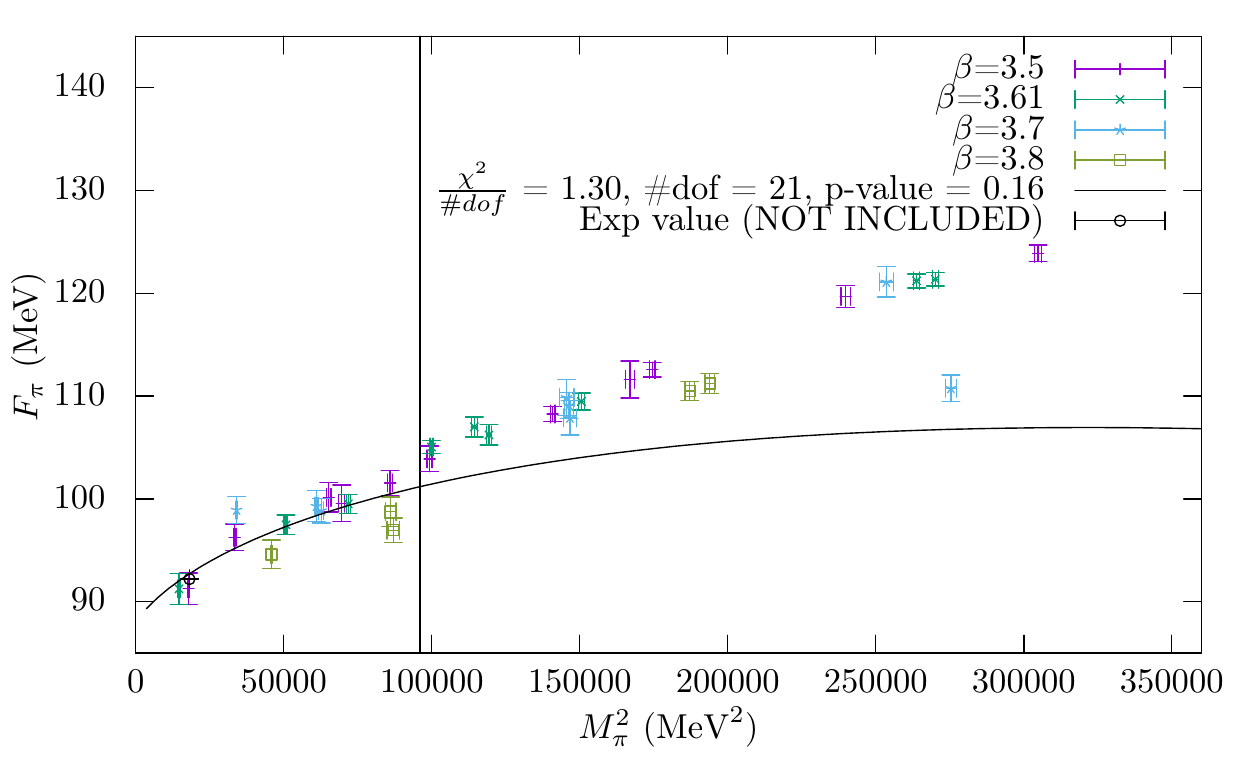}\\
\includegraphics[width=0.5\textwidth]{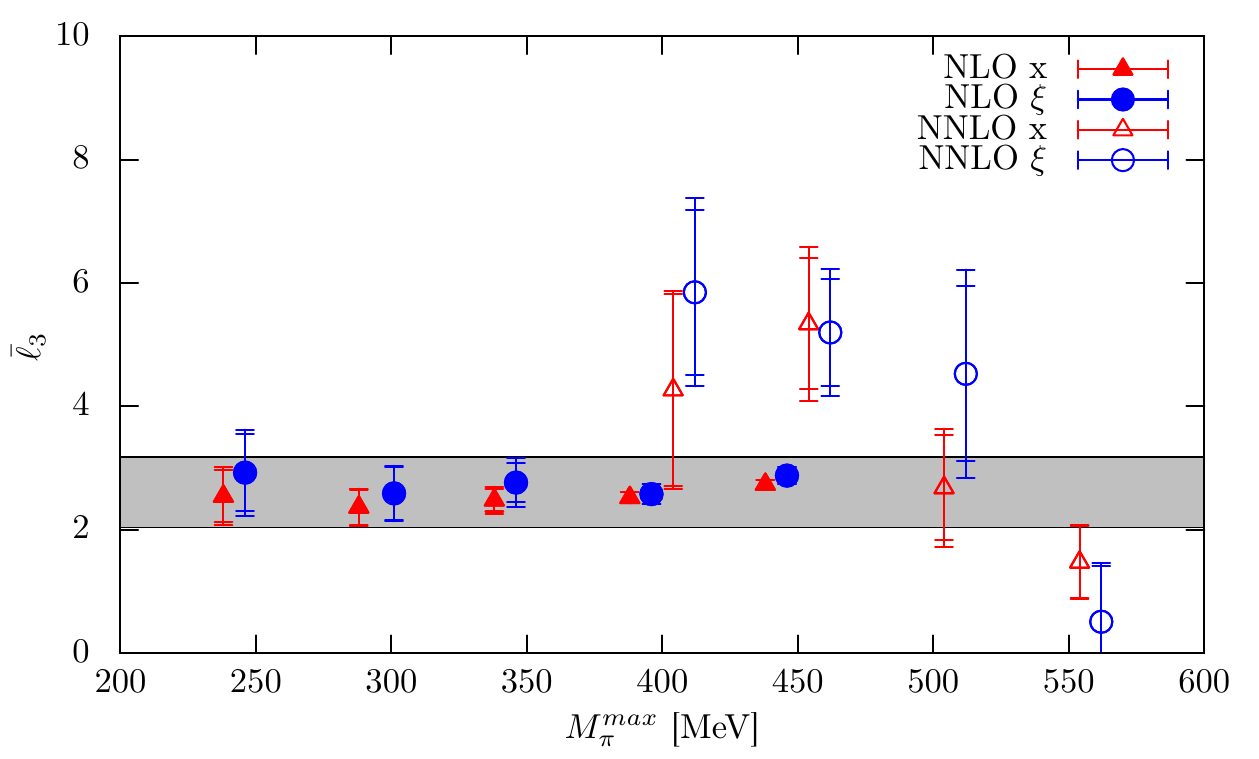}%
\includegraphics[width=0.5\textwidth]{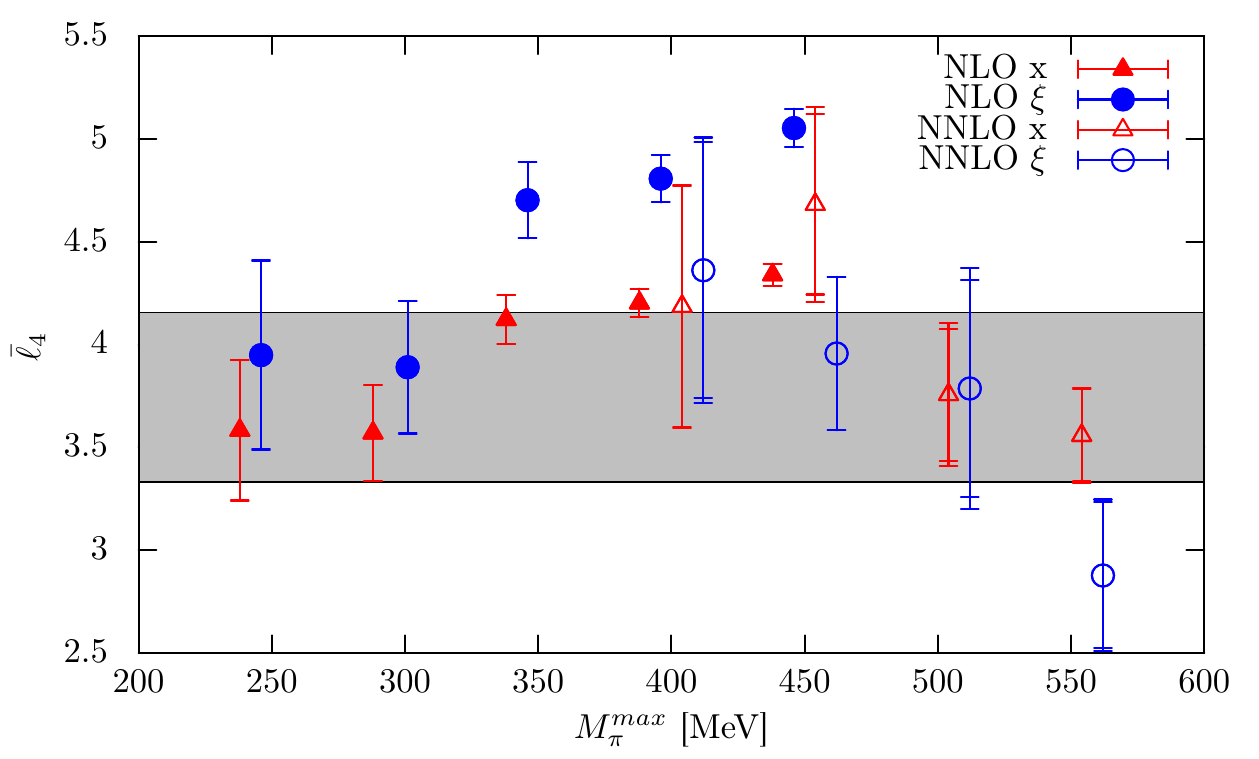}%
\vspace*{-2mm}
\caption{\label{fig:wils_NLO_xi}
Top: $\xi$ expansion fit at NLO of $m_{ud}/\Mpi^2$ and $\Fpi$ versus $\Mpi^2$
with $\Mpi^\mr{max}=300\MeV$.
Bottom: fit parameters $\bar\ell_3,\bar\ell_4$ as a function of $\Mpi^\mr{max}$.}
\end{figure}

We perform joint LO+NLO fits, this time in the $x$ and $\xi$ expansion, and
monitor how parameters shift as a function of $\Mpi^\mr{max}$.
Fig.\,\ref{fig:wils_NLO_x} shows our results for the LECs at the LO;
$B$ is fairly insensitive to this cut, while $F$ is significantly affected if
$\Mpi^\mr{max}>300\MeV$.
Fig.\,\ref{fig:wils_NLO_xi} shows our results for the LECs at the NLO;
$\bar\ell_3$ is fairly robust, while $\bar\ell_4$ shows a clear trend.
All these findings are in complete analogy to what was found in the
staggered case.
A new ingredient is that we can now compare the $x$ and the $\xi$ expansion results
(red triangles versus blue bullets) in case there is a drift.
For $F$ there is no discrepancy (last panel of Fig.\,\ref{fig:wils_NLO_x}), while
for $\bar\ell_4$ the onset of a discrepancy seems to signal the end of the
regime where NLO ChPT is applicable (last panel of Fig.\,\ref{fig:wils_NLO_xi}).


\subsection{NNLO fit via $x$ and $\xi$ expansion}

\begin{figure}
\includegraphics[width=0.5\textwidth]{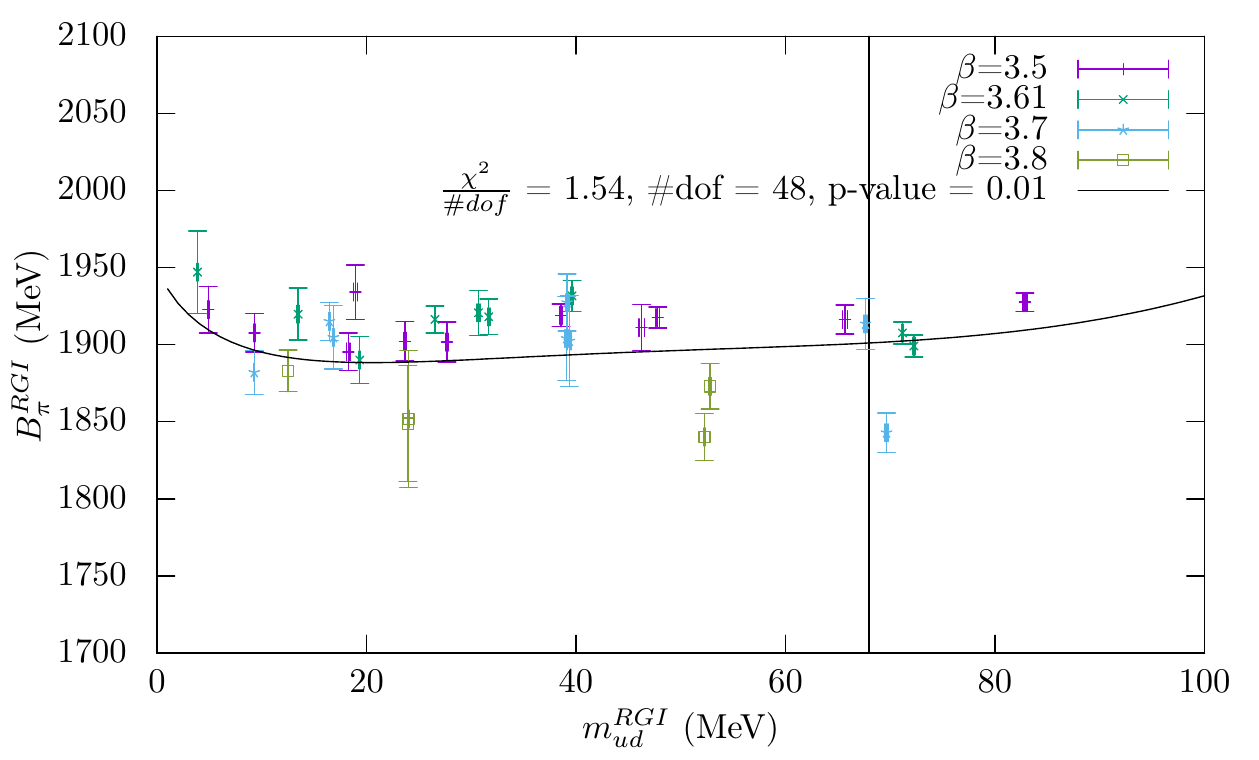}%
\includegraphics[width=0.5\textwidth]{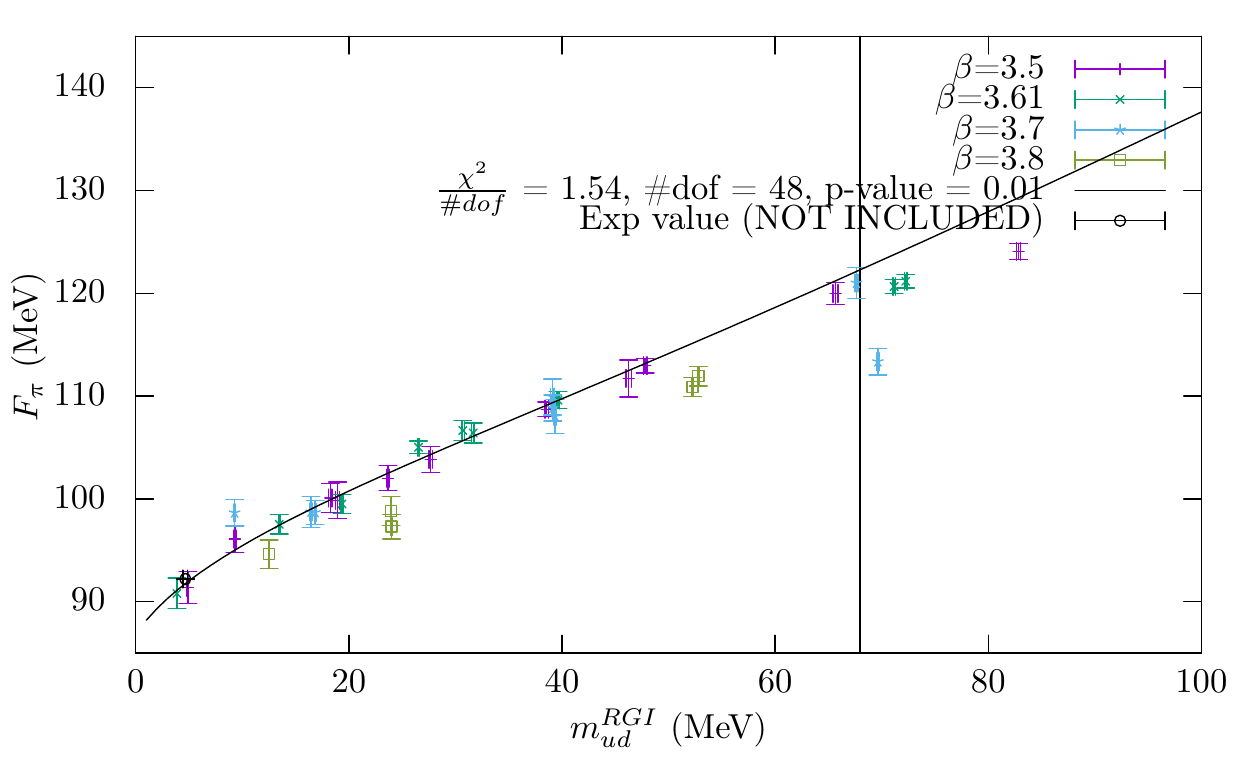}\\
\includegraphics[width=0.5\textwidth]{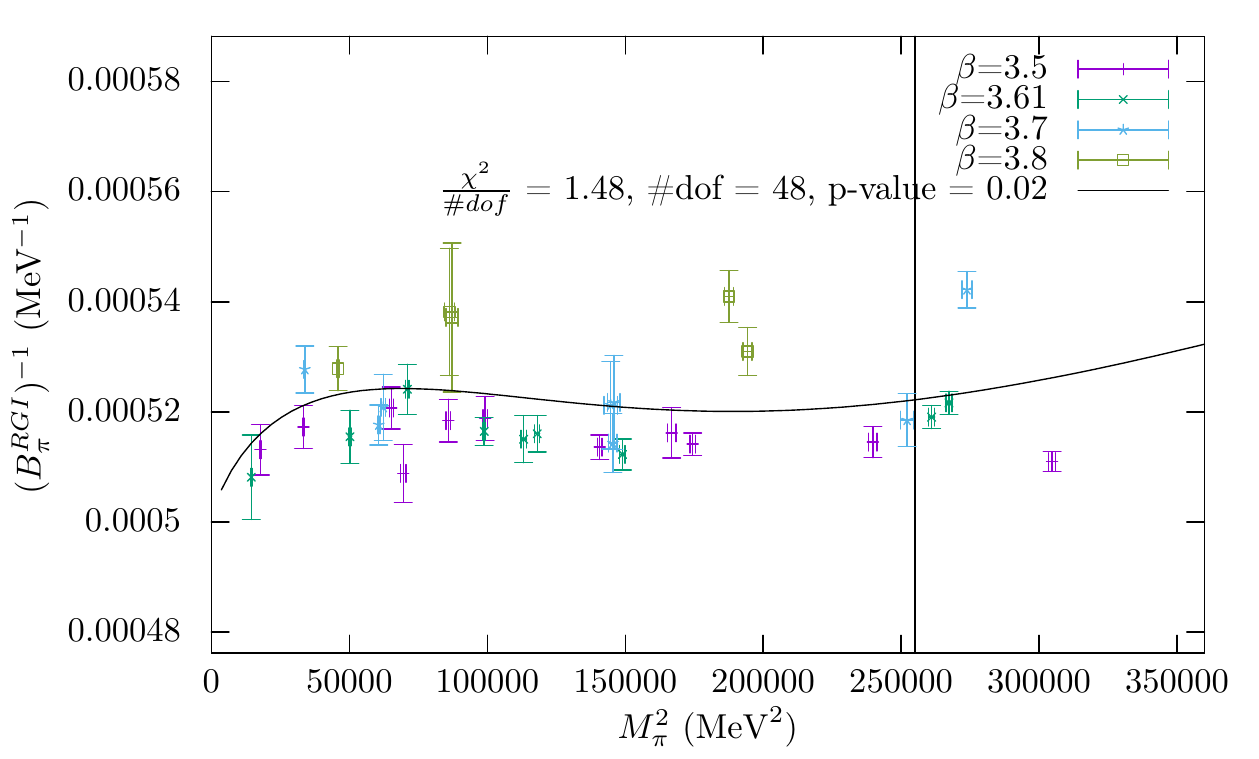}%
\includegraphics[width=0.5\textwidth]{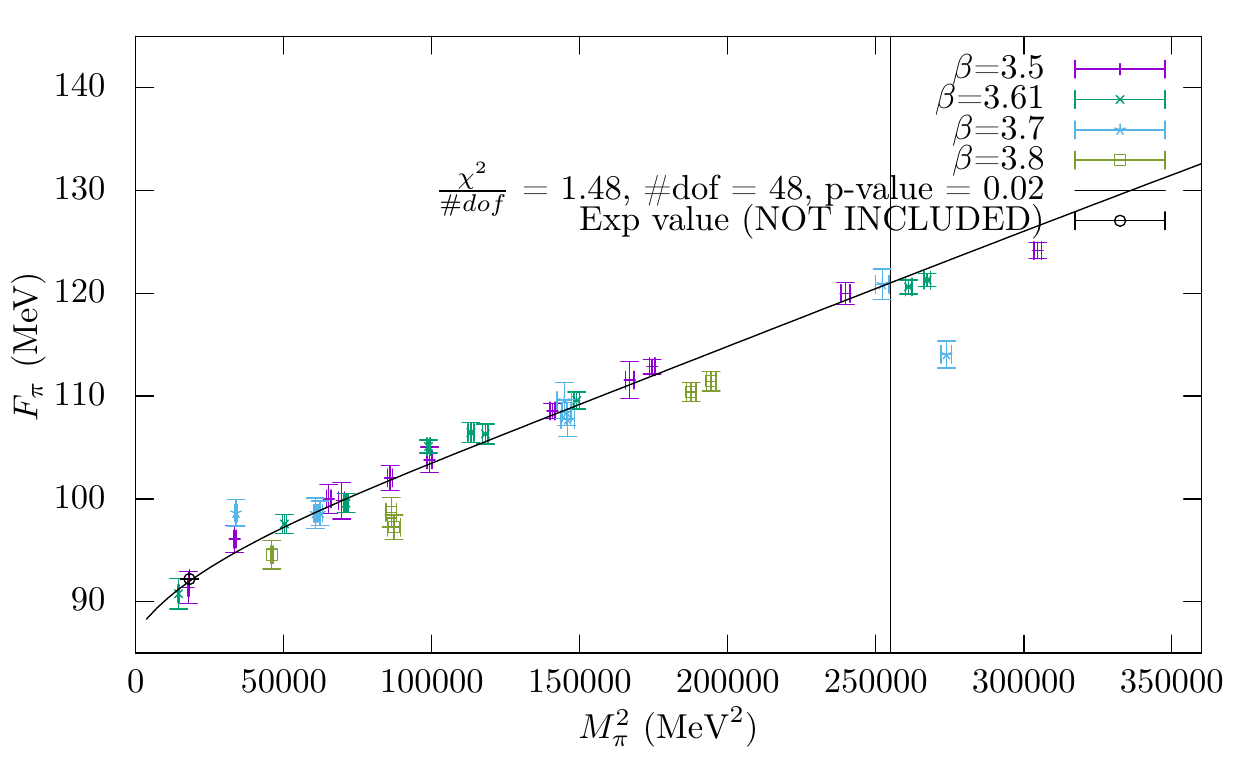}%
\vspace*{-2mm}
\caption{\label{fig:wils_NNLO}
LO+NLO+NNLO fit in $x$ expansion (top) or $\xi$ expansion (bottom) of $\Mpi^2/m_{ud}$
or $m_{ud}/\Mpi^2$ (left) and $\Fpi$ (right) versus $m_{ud}$ or $\Mpi^2$, respectively.}
\end{figure}

\begin{figure}
\includegraphics[width=0.5\textwidth]{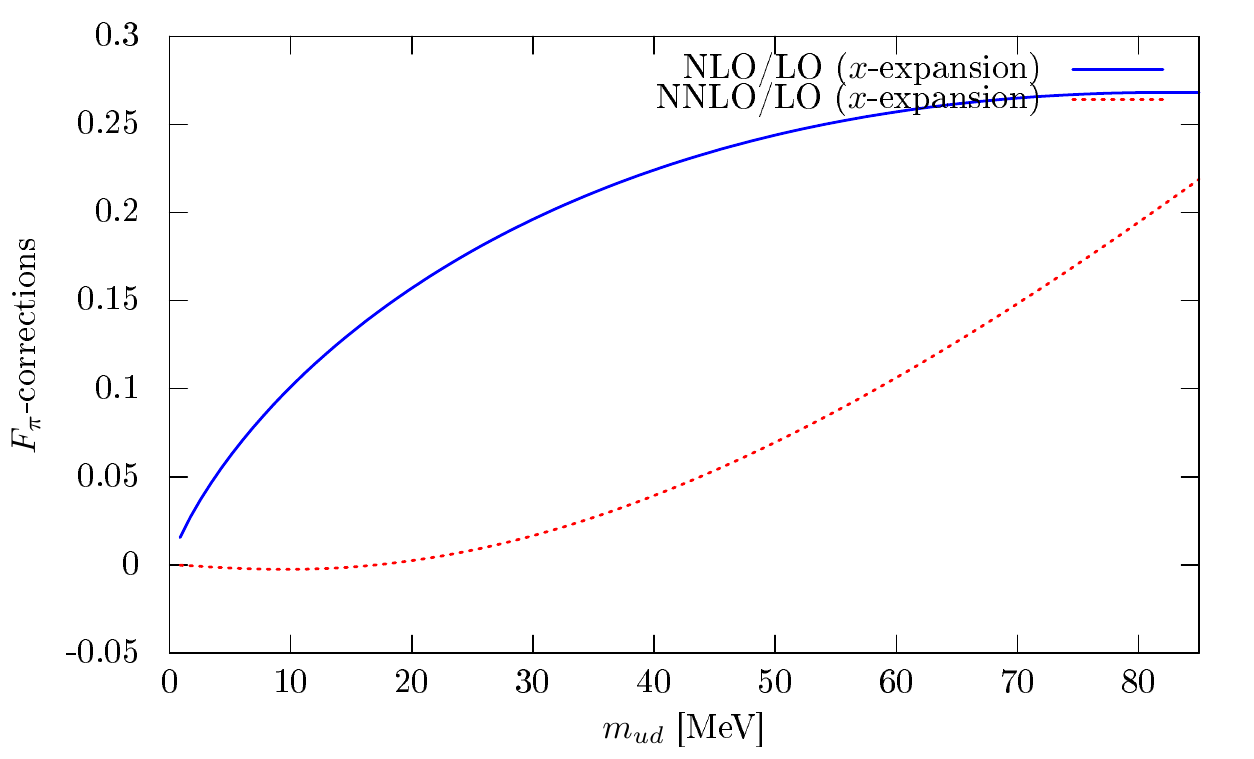}%
\includegraphics[width=0.5\textwidth]{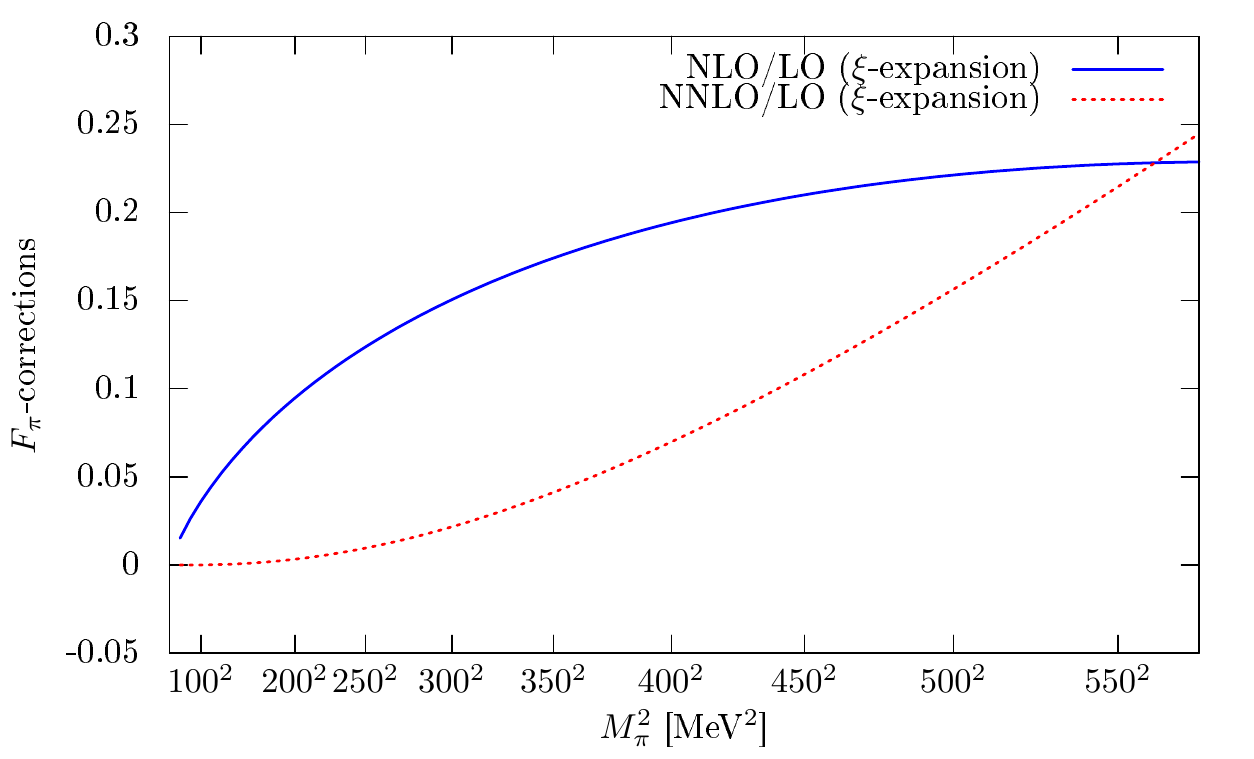}%
\vspace*{-2mm}
\caption{\label{fig:wils_breakup}
Breakup of LO+NLO+NNLO fit in $x$ or $\xi$ expansion to give NLO/LO and NNLO/LO ratios.}
\end{figure}

Given the large dataset, we attempt a provisional LO+NLO+NNLO fit.
It is clear that such fits necessitate the inclusion of somewhat higher
$\Mpi^\mr{max}$ values, see Fig.\,\ref{fig:wils_NNLO}.
We are not interested in the LECs at the NNLO, but rather how the LO and NLO
counterparts compare to those obtained from the direct LO+NLO fit.
The open symbols in Fig.\,\ref{fig:wils_NLO_x} and Fig.\,\ref{fig:wils_NLO_xi}
indicate that they tend to come with larger statistical errors, but within
errors they are reasonably consistent with the earlier results.

Once more we can break up the complete fit into its LO, NLO, and NNLO
contributions, and compare their relative importance.
From Fig.\,\ref{fig:wils_breakup} we learn that in $\Fpi=\Fpi(m_{ud})$ or
$\Fpi=\Fpi(\Mpi^2)$ the NLO contribution stays saturated at about 25\% of the
LO contribution, but the NNLO/NLO ratio exceeds $\frac{1}{2}$ at about
$\Mpi\sim450\MeV$ (which is a first sign of distress), and exceeds $1$ at
about $\Mpi\sim550\MeV$ (which clearly signals the breakdown of the chiral
expansion).
These results are again in qualitative agreement with what was found in the
staggered case.


\subsection{Sensitivity of LO+NLO fit on pruning data from below}

\begin{figure}
\includegraphics[width=0.5\textwidth]{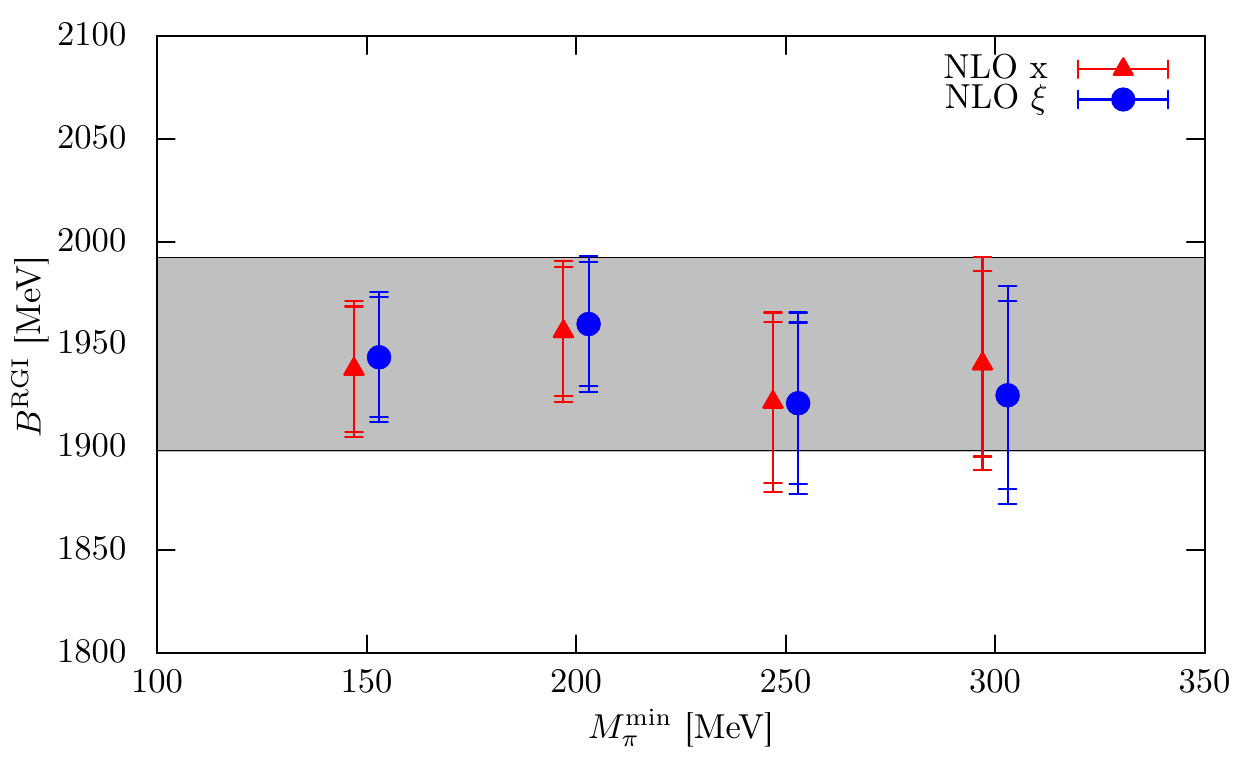}%
\includegraphics[width=0.5\textwidth]{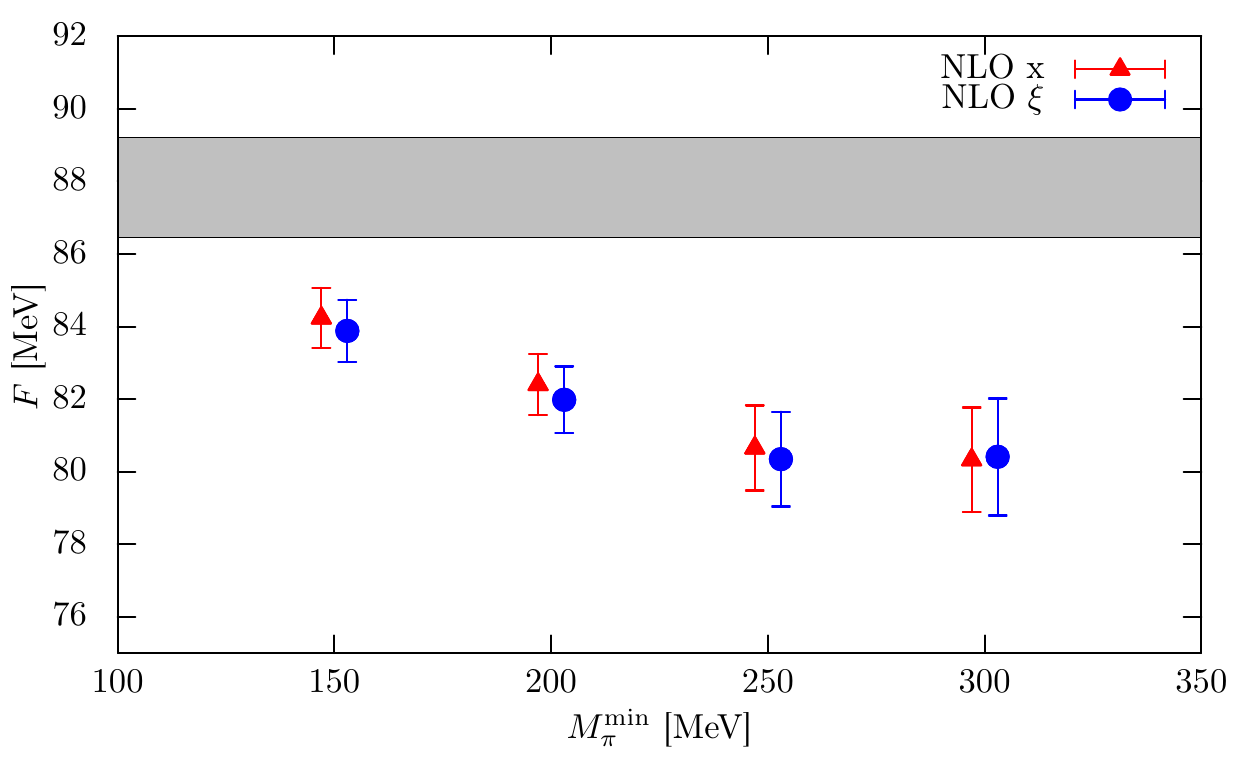}\\
\includegraphics[width=0.5\textwidth]{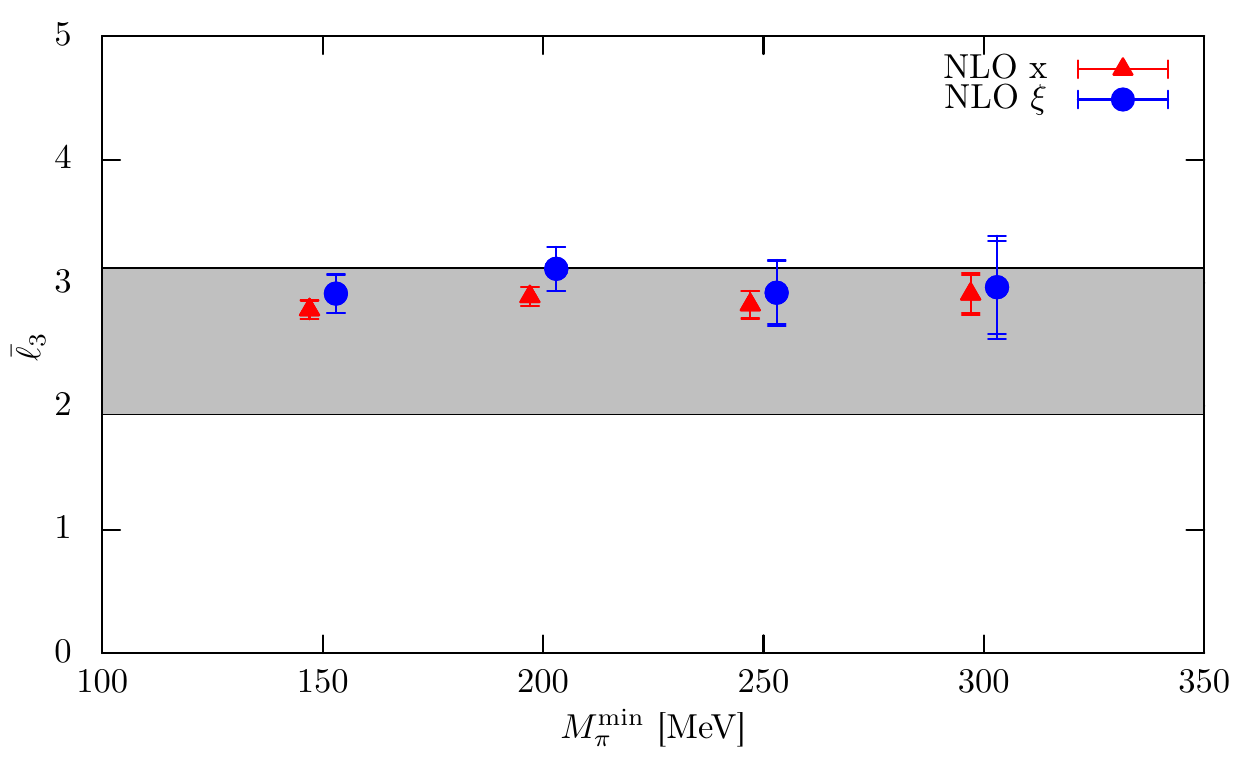}%
\includegraphics[width=0.5\textwidth]{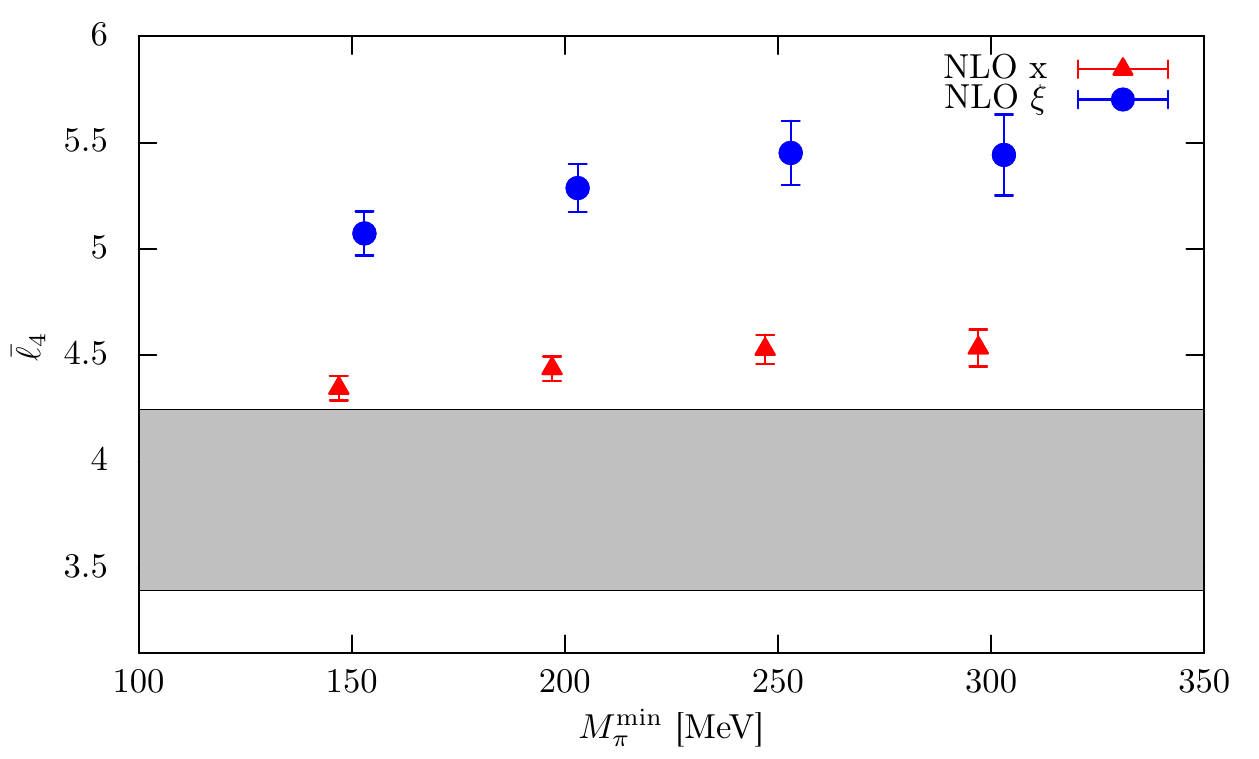}%
\vspace*{-2mm}
\caption{\label{fig:wils_extracuts}
Sensitivity of LO (top) and NLO (bottom) low-energy constants on $\Mpi^\mr{min}$
in LO+NLO fits.}
\end{figure}

Fig.\,\ref{fig:wils_extracuts} shows the results for $B,F$ (top) as well
as $\bar\ell_3,\bar\ell_4$ (bottom) as a function of $\Mpi^\mr{min}$.
We find that $B,\bar\ell_3$ are fairly robust, while $F,\bar\ell_4$
tend to come out too low and too high, respectively, in view of results that
include our more chiral data (indicated by the gray bands).
Again, the discrepancy between $x$ and $\xi$ expansion may signal an
inappropriate mass range ($\bar\ell_4$) but need not do so ($F$).


\section{FLAG review of LECs in SU(2) and SU(3) ChPT \label{sec:flag}}


Though this is a topical contribution, it might be adequate to highlight the
FLAG summary \cite{Aoki:2013ldr} of LECs to show that there are several fine
lattice calculations of LECs in SU(2) and SU(3) ChPT.


\subsection{Summary of SU(2) LECs at NLO}

\begin{figure}
\includegraphics[width=0.48\textwidth]{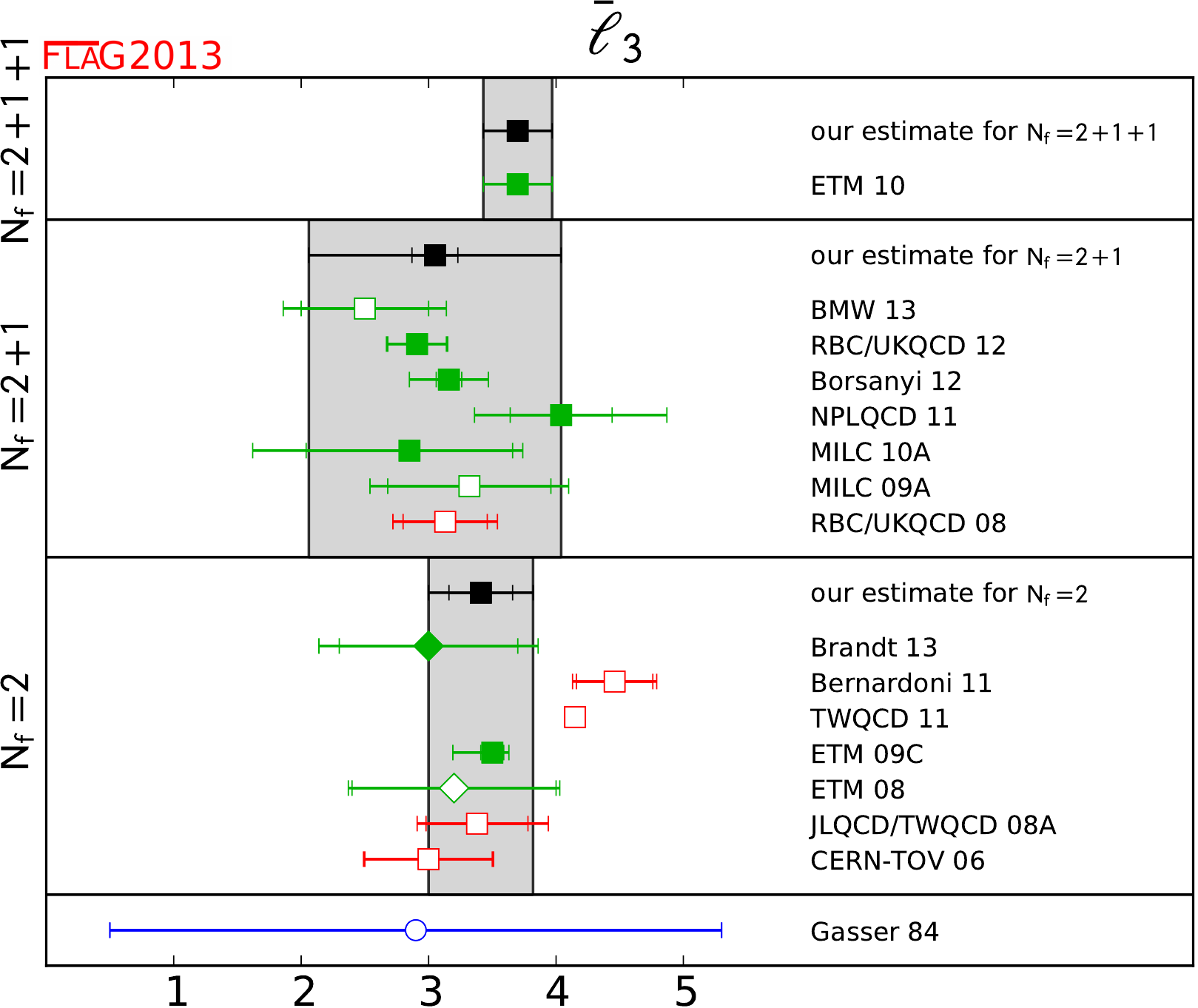}\hfill
\includegraphics[width=0.48\textwidth]{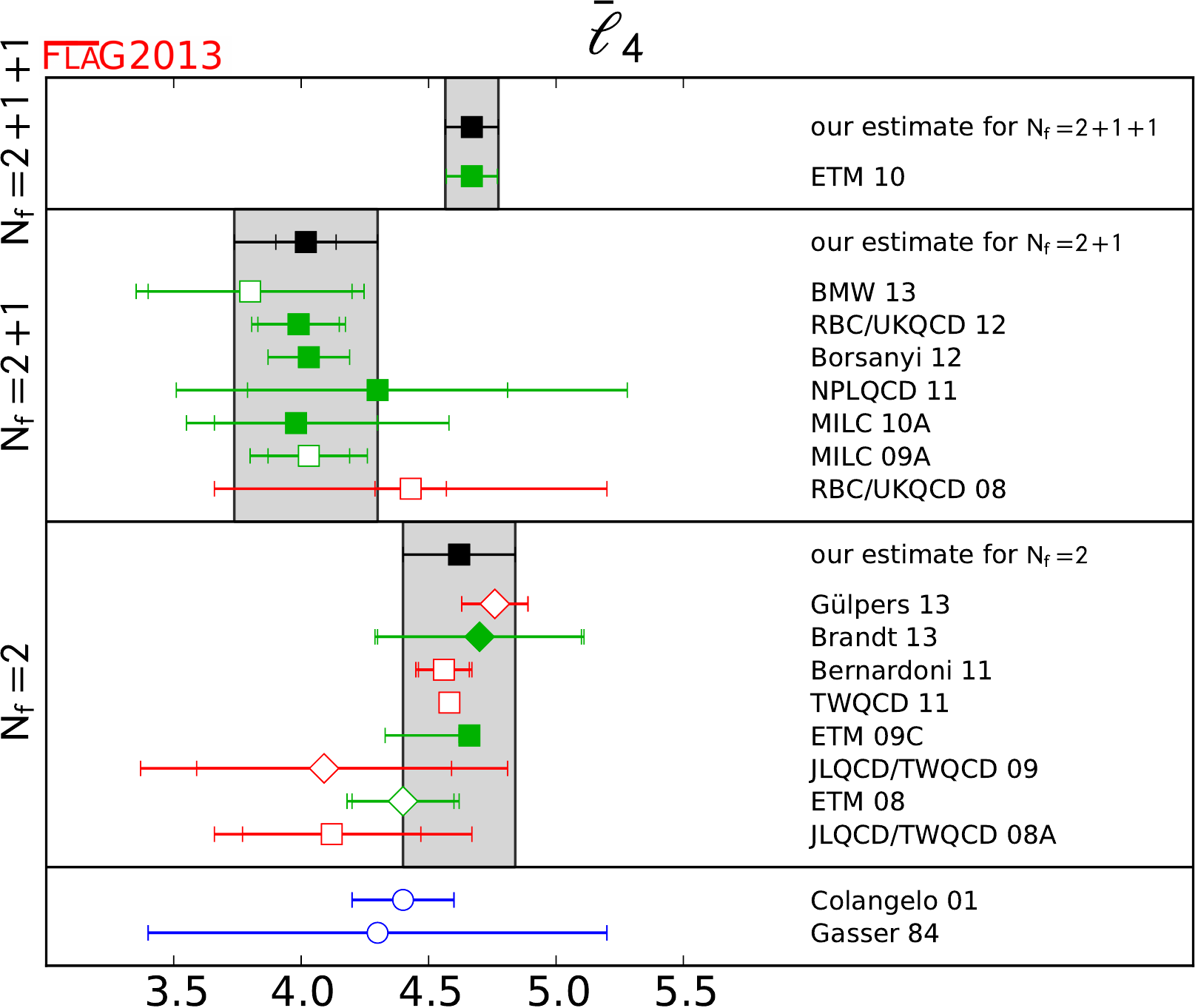}%
\vspace*{-2mm}
\caption{\label{fig:flag_su2}
Summary of the SU(2) NLO low-energy constants $\bar\ell_3$ and $\bar\ell_4$
as compiled by FLAG \cite{Aoki:2013ldr}.}
\end{figure}

Fig.\,\ref{fig:flag_su2} shows the FLAG compilation of SU(2) LECs at the NLO.
The results are grouped into the categories $\Nf=2$, $\Nf=2+1$, and
$\Nf=2+1+1$.
In addition, one or two phenomenological calculations of high standing are
added for comparison.
Some of the results are shown with red symbols, because one ingredient of the
calculation is not state-of-the-art (e.g.\ data do not probe the chiral regime,
or just one lattice spacing is used).
Results which passed such quality checks are represented by green symbols.
A filled symbol indicates that the result did enter the FLAG recommended
value, an open symbol means that it did not (e.g.\ because the paper is not
yet published or because it has been superseded by a more recent result by
the same collaboration).

FLAG aims for very conservative error estimates.
A standard mathematical average is formed, but then the error is stretched until
the gray $\pm1\sigma$ band covers the central values of all calculations which
entered.
This is done for each $\Nf$ so that one can observe a potential $\Nf$ dependence.

Concerning $\bar\ell_3$ the lattice approach is a success story.
The results from $\Nf=2$, $\Nf=2+1$ and $\Nf=2+1+1$ studies are consistent,
and they achieve better precision than the phenomenological estimate
``Gasser 84''.
For $\bar\ell_4$ the situation is more challenging.
For each $\Nf$ the lattice results seem consistent, but there is a remnant
dependence on whether a strange and/or charm quark is included in the sea.
In principle, such an $\Nf$ dependence is possible, but one would expect it
to be monotonic.
In addition, the lattice has difficulties in beating ``Colangelo 01'' in terms
of precision.


\subsection{Summary of SU(3) LECs at NLO}

\begin{figure}
\includegraphics[width=0.48\textwidth]{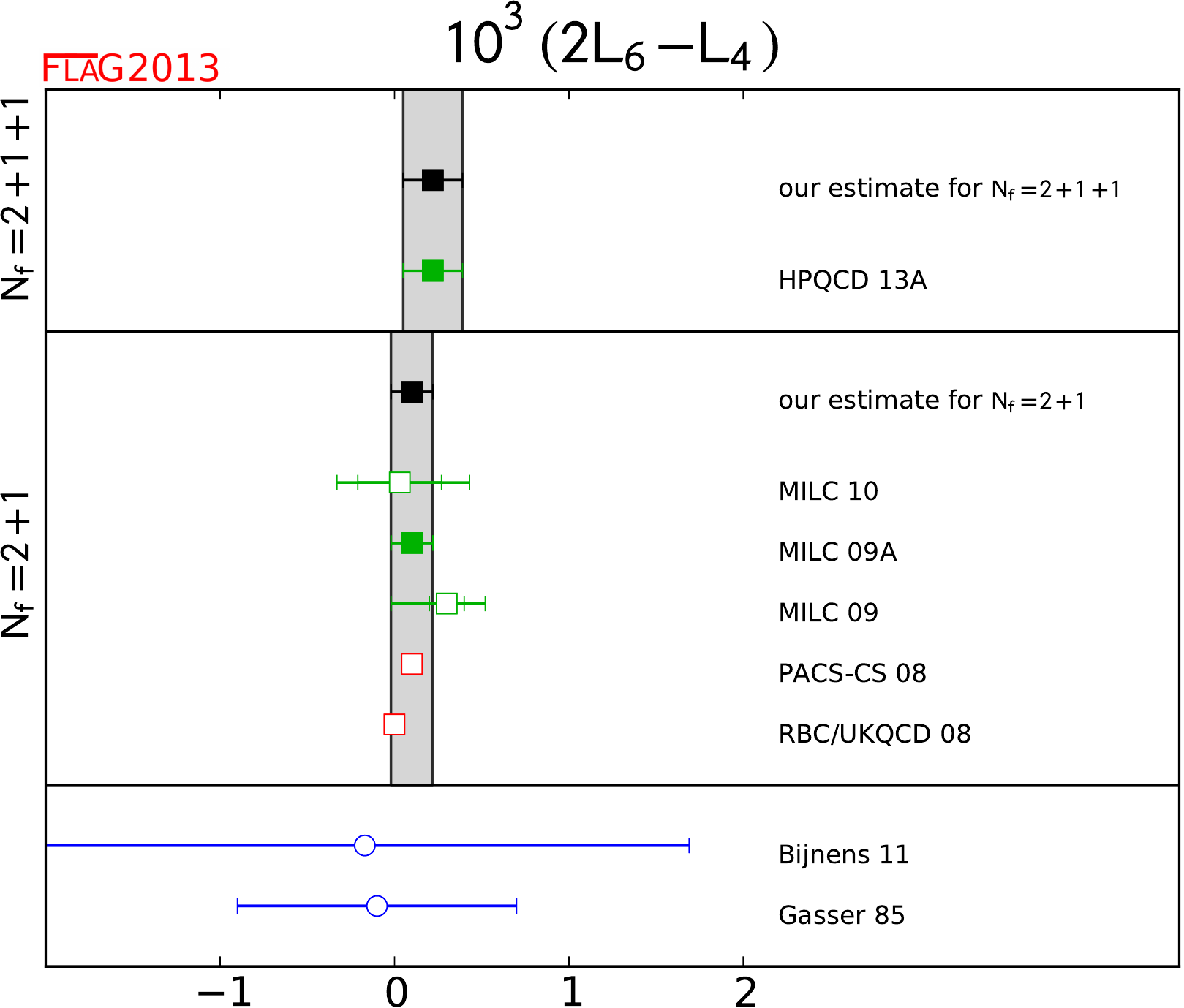}\hfill
\includegraphics[width=0.48\textwidth]{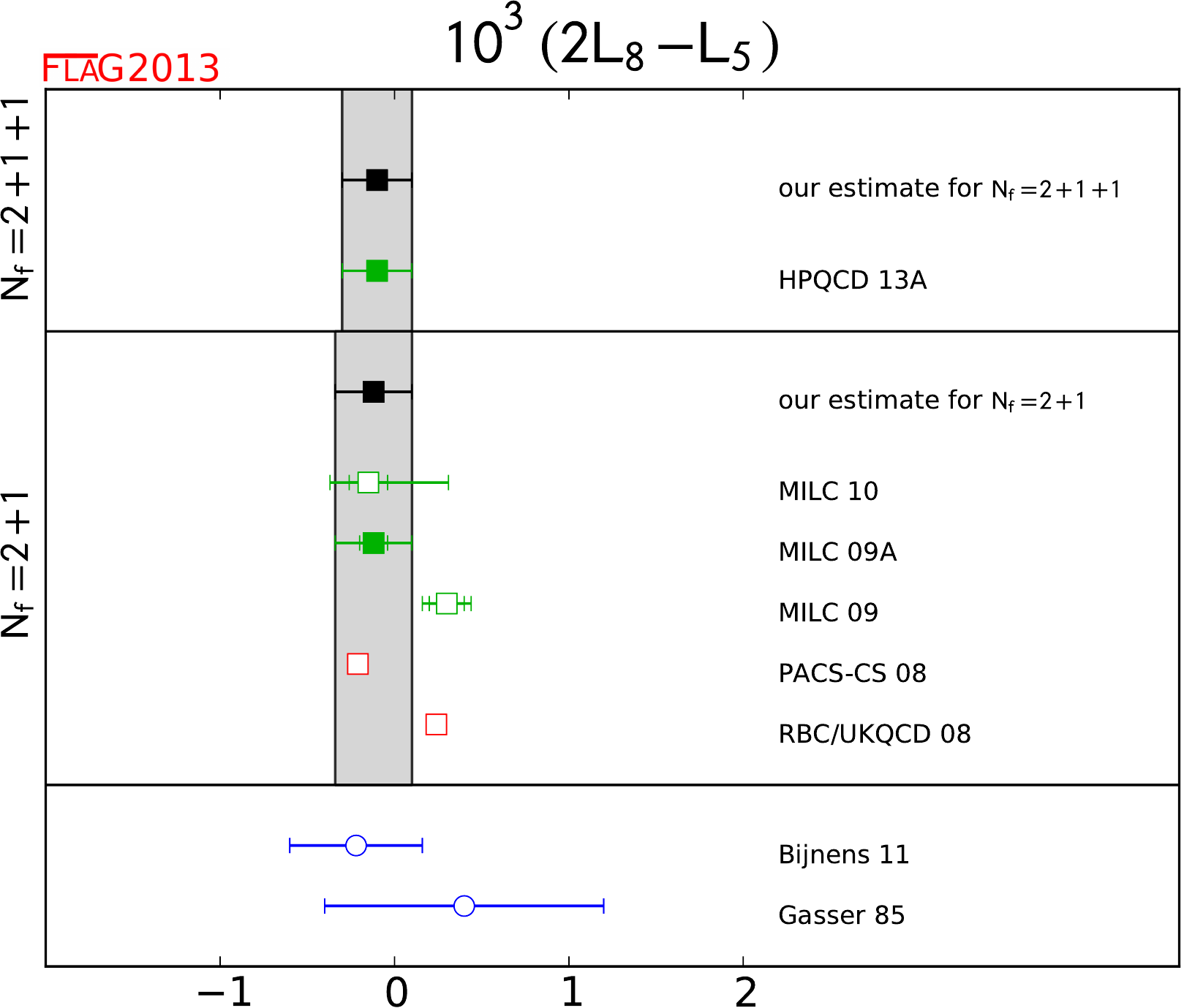}%
\vspace*{-2mm}
\caption{\label{fig:flag_su3}
Summary of the SU(3) NLO low-energy constant combinations
$2L_6^\mr{ren}-L_4^\mr{ren}$ and $2L_8^\mr{ren}-L_5^\mr{ren}$ as compiled by
FLAG \cite{Aoki:2013ldr}. The renormalization scale is $\mu\sim770\MeV$, as
is customary in phenomenology.}
\end{figure}

Fig.\,\ref{fig:flag_su3} shows the FLAG compilation of SU(3) LECs at the NLO.
The results are grouped into the categories $\Nf=2+1$ and $\Nf=2+1+1$.
Again, one or two phenomenological calculations of high standing are
added for comparison.
The main difference to the SU(2) case is that there are fewer lattice
determinations.
This is not so much an effect of few collaborations generating $\Nf=2+1$
ensembles, but of the requirement that, in order to control the SU(3)
expansion, some ensembles with $m_s\ll m_s^\mr{phys}$ must be available
(compare the discussion in Sec.\,\ref{sec:intro} and Subsec.\,\ref{sub:su2vssu3}).

The situation looks quite favorable for the lattice approach, for both $2L_6-L_4$
and $2L_8-L_5$.
The lattice results seem reasonably consistent, there is no visible $\Nf$ dependence,
and the latest generation of results is significantly more precise than the best
phenomenological determinations.

In the future one would like to see tests of the large-$\Nc$ prediction $L_{4,6}\to0$,
and one would like to see precision results for the flavor breaking ratios
$F^{(2)}/F^{(3)}$, $\Sigma^{(2)}/\Sigma^{(3)}$, $B^{(2)}/B^{(3)}$,
in order to test the Zweig rule (see Ref.\,\cite{Aoki:2013ldr} for details).
As discussed in Subsec.\,\ref{sub:su2vssu3} such studies require two different
chiral limits to be performed from one set of $\Nf=2+1$ or $\Nf=2+1+1$ data.


\section{Assorted remarks}


Let me finish this proceedings contribution with two brief remarks -- one
concerning the way how LECs are calculated, one concerning the impact of a growing
number of \emph{light} flavors.


\subsection{Rationale for log-free compounds}

\begin{figure}
\hspace*{-10pt}%
\includegraphics[height=57mm]{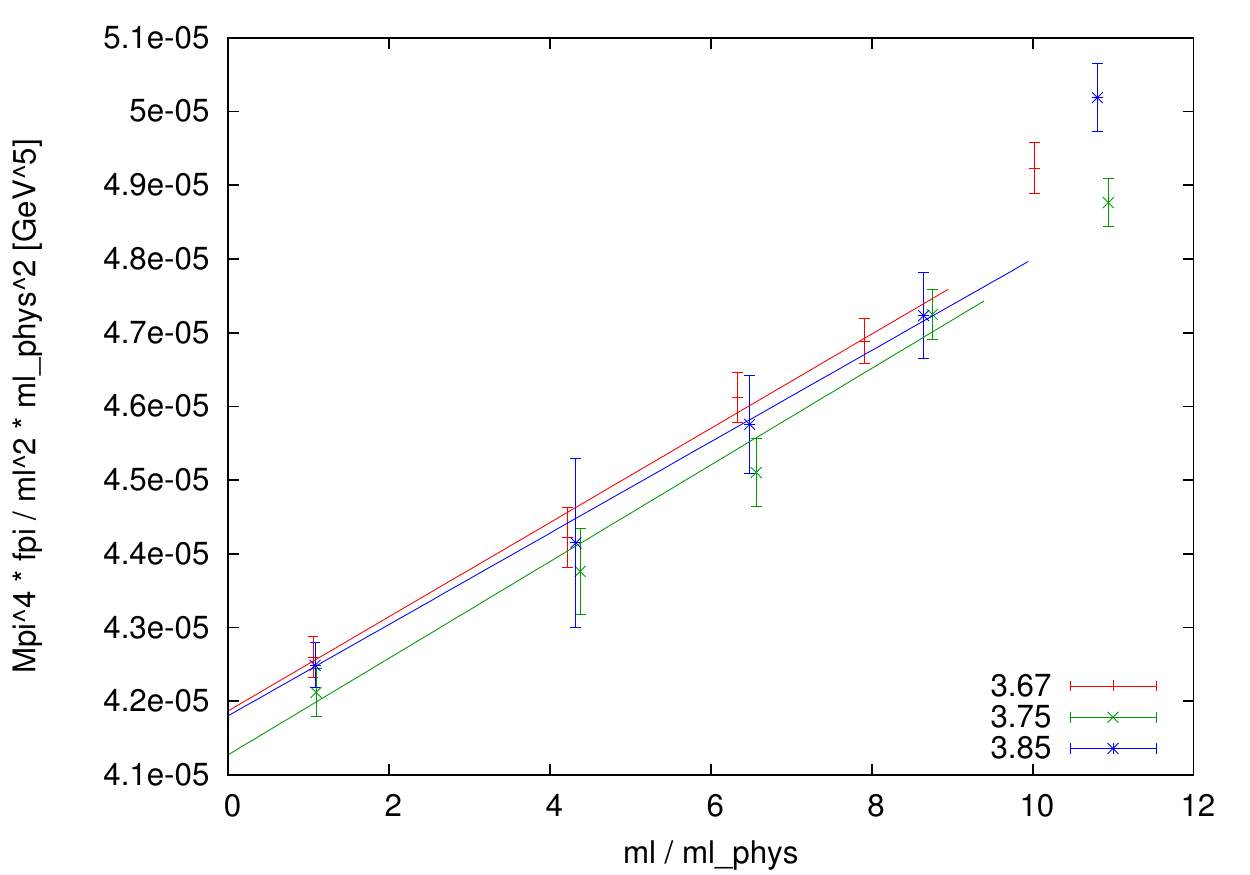}%
\includegraphics[height=55mm]{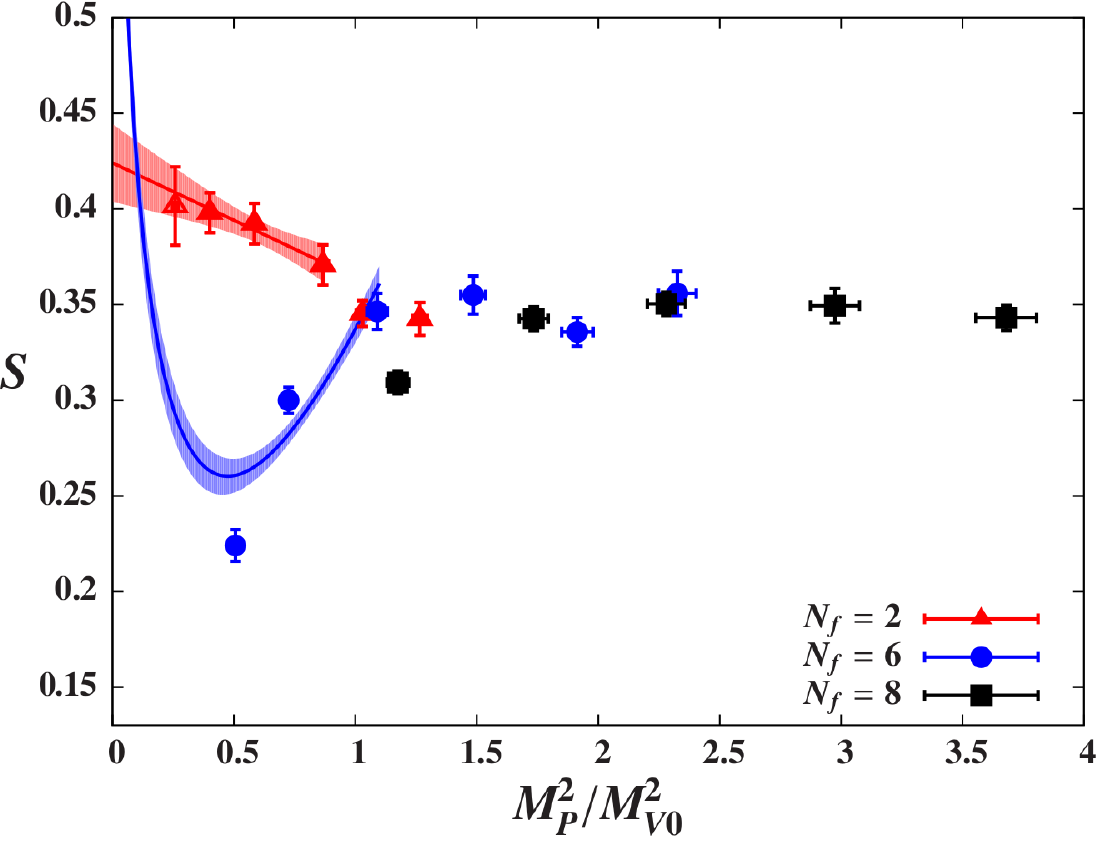}%
\vspace*{-2mm}
\caption{\label{fig:remarks}
Left: $\Mpi^4\fpi^{}/m_{ud}^2$ as a function of $m_{ud}$, with quark masses
normalized by $m_{ud}^\mr{phys}$, at the three finest lattice spacings of
Ref.\,\cite{Borsanyi:2012zv}. Right: $S$-parameter for $\Nf=2,6,8$ as a function
of $M_P^2/M_V^2$ from Ref.\,\cite{Appelquist:2014zsa}.}
\end{figure}

As discussed in Sec.\,\ref{sec:stag} and Sec.\,\ref{sec:wils} the dominant source
of systematic uncertainty in a lattice determination of ChPT LECs at the NLO is
typically the uncertainty about the impact that the choice of the fitting range
$[\Mpi^\mr{min},\Mpi^\mr{max}]$ has.
It was argued that sufficiently fine grained and sufficiently precise data in the
range $[135\MeV,350\MeV]$ are sufficient to determine the LO LECs $F,B$ as
well as the NLO LECs $\bar\ell_3,\bar\ell_4$ of SU(2) ChPT.
The main difficulty is the lack of a clear criterion to decide where a standard
chiral logarithm $\propto\Mpi^2\log(\Mpi^2/\Lambda^2)$ is present and where
something else, e.g.\ a higher-order chiral log or strong cut-off effects,
contributes significantly.

For certain linear combinations of LECs the task is considerably easier.
For instance
\beq
\Mpi^4\Fpi=M^4F\;
\big\{
1+x\log(\Lambda_4^2/\Lambda_3^2)+O(x^2)
\big\}
\label{logfree}
\eeq
is a direct consequence of (\ref{chpt_Mpi},\,\ref{chpt_Fpi}).
In other words, the ratio $\Lambda_4^2/\Lambda_3^2$ or equivalently the difference
$\bar\ell_4-\bar\ell_3=\log(\Lambda_4^2/\Lambda_3^2)$ can be determined from the
behavior of $\Mpi^4\Fpi$ as a function of the quark mass in the regime where
this behavior is \emph{linear}.
The left panel of Fig.\,\ref{fig:remarks} shows this behavior for the three finest
lattice spacings of Ref.\,\cite{Borsanyi:2012zv} up to $m_{ud}=9m_{ud}^\mr{phys}$
or $\Mpi\sim400\MeV$.
Relative to formula (\ref{logfree}) there is an extra factor
$\sqrt{2}(m_{ud}^\mr{phys}/m_{ud})^2$, hence the intercept determines
$(2Bm_{ud}^\mr{phys})^2f$, and the slope $6.4(4)(4)\,10^{-7}$ in terms of
$m_{ud}/m_{ud}^\mr{phys}$ determines
$\ch_\mr{phys}^3/(8\pi^2 f)\cdot\log(\Lambda_4^2/\Lambda_3^2)$.
With $\ch_\mr{phys},f$ from that work this yields
$\bar\ell_4-\bar\ell_3=0.96(06)(13)$ which is perfectly consistent with
Ref.\,\cite{Borsanyi:2012zv}.
Note, finally, that the combination $\Mpi^4\Fpi$ is free of finite-volume effects
through NLO.


The flip-side is that the orthogonal combination
$\bar\ell_4+\bar\ell_3$ is determined from $\Mpi^4/(\Fpi m_{ud})$, and
this combination is pounded with both genuine chiral logs and strong finite
volume effects.
Finally, let me add that the combination (\ref{logfree}) follows from
$\gamma_3=-\frac{1}{2},\gamma_4=2$ in SU(2) ChPT \cite{Gasser:1983yg}.
The replication of this trick in SU(3) ChPT is straightforward,
since the $\Gamma_i$ are known \cite{Gasser:1984gg}.


\subsection{$S$-parameter in $\Nf\!=\!6,8,...$ theories}

QCD is easily generalized to $\Nc$ colors and $\Nf$ light flavors.
That technical difficulties increase sharply with $\Nf$, at fixed $\Nc=3$, is
suggested by the right panel of Fig.\,\ref{fig:remarks}, which is taken from
Ref.\,\cite{Appelquist:2014zsa}.
The motivation to explore candidates of EW symmetry breaking is of no concern
for us.
What matters is that they study an observable, the $S$-parameter, for which
there is a chiral prediction (red and blue fits).
For $\Nf=2$ there seems to be good agreement.
For $\Nf=6$ it seems to become difficult to enter the mass regime where the
chiral expansion is applicable.
For $\Nf=8$ the authors do not even attempt a chiral fit.
According to conventional wisdom cut-off effects are unlikely to cause these
difficulties, since the authors use domain-wall fermions with
$am_\mr{res}\simeq0.003$.


\section{Summary \label{sec:summary}}


In summary it seems fair to say that the computation of LO and NLO LECs of
both SU(2) and SU(3) ChPT has grown into a mature field.
In general there are two types of LECs -- those which parameterize the momentum
dependence of QCD Green's function at low energy, and those which parameterize
the quark mass dependence.
The former set of LECs is usually well determined from experiment, but for the
latter set of LECs the lattice has the unique opportunity of varying the quark
mass.
Accordingly, the initial statement in this paragraph is meant for this latter
category.

These LECs are determined from matching some lattice data to LO+NLO
chiral formulas.
This implies that beyond the standard sources of systematic uncertainty
($a\!\to\!0$, $L\!\to\!\infty$, interpolation/extrapolation to $m_q^\mr{phys}$)
the mass range used in the fit is an additional source of systematics, and in
many cases it turns out to be the dominant one.
In this proceedings contribution evidence was provided that in case of the
low-energy constants $\bar\ell_3,\bar\ell_4$ (or $\Lambda_3,\Lambda_4$) these
systematic uncertainties can be controlled and reliably estimated, if
sufficiently precise data between 135\,MeV and about 350\,MeV are available.
The preferred fits use a mass range up to 240\,MeV and 300\,MeV for staggered
and Wilson fermions, respectively, but alternative (lower and higher)
$\Mpi^\mr{max}$ values are needed to assess to systematic uncertainty.
In addition, it was found that an $\Mpi^\mr{min}$ value not far from 135\,MeV
is needed to obtain correct results.

In some cases exploratory LO+NLO+NNLO fits have been attempted, and it is
reassuring to see that the LO and NLO coefficients determined in this way
agree with those obtained from direct LO+NLO fits.
A break-up of these fits reveals some degradation of the convergence of the
chiral expansion near $\sim\!400$\,MeV and suggests a complete breakdown
beyond $\sim\!500$\,MeV.
Also a comparison of results from LO+NLO fits in the $x$ and $\xi$ expansion
can be useful to detect some stress in the chiral series, but it need not
always provide this kind of service.

Last but not least let me recall that the field of pseudo-Goldstone boson
dynamics in standard QCD is particularly favorable to ChPT.
Evidence is mounting that it gets progressively harder to enter the
chiral regime if $\Nf$ is increased, and of course we know since a long
time that even at $\Nf\!=\!2$ the chiral expansion converges more slowly in
the nucleon sector.
It remains a noble goal to investigate whether this is linked to a change
of the role played by scalar resonances.

\bigskip
\bigskip

\noindent
{\bf Acknowledgments}:
I am indebted to my colleagues in the BMW collaboration and in FLAG.
Special thanks go to Enno Scholz and Alfonso Sastre for the fine analysis
work that was essential for Refs.\,\cite{Borsanyi:2012zv,Durr:2013goa}.
This work is in part supported by the German DFG through SFB-TR-55.


\end{document}